\documentclass[12pt]{article}
\usepackage[utf8]{inputenc}
\usepackage[margin=1in]{geometry}
\usepackage{bm}
\usepackage{amsmath, color, natbib}
\usepackage{amssymb}
\usepackage{amsthm}
\usepackage{graphicx}
\usepackage{amsfonts}
\usepackage{mathrsfs}
\numberwithin{equation}{section}
\usepackage{booktabs}
\usepackage{caption}
\usepackage{subcaption}
\usepackage{float}
\usepackage{adjustbox, url}
\usepackage[table]{xcolor}
\usepackage[doublespacing]{setspace}
\usepackage{authblk}
\usepackage{multirow}
\usepackage{soul}
\setstcolor{blue}

\newcommand{\bX}{\bm{X}}
\newcommand{\bth}{\bm{\theta}}
\newcommand{\bbeta}{\bm{\beta}}
\newcommand{\bgamma}{\bm{\gamma}}
\newcommand{\tbeta}{\tilde{\beta}}

\newcommand{\btth}{\tilde{\bm{\theta}}}
\newcommand{\btbeta}{\tilde{\bm{\beta}}}
\newcommand{\btgamma}{\tilde{\bm{\gamma}}}
\newcommand{\mT}{\mathrm{T}}
\newcommand{\pkg}[1]{{\fontseries{b}\selectfont #1}}
\newcolumntype{P}[1]{>{\centering\arraybackslash}p{#1}}

\title{A Bayesian Nonparametric Approach for Inferring Drug \mbox{Combination Effects on Mental Health in People with HIV}}

\author[1]{Wei Jin}
\author[2]{Yang Ni}
\author[3, 4]{Leah H. Rubin}
\author[5]{Amanda B. Spence}
\author[1, *]{\mbox{Yanxun Xu}}
\affil[1]{Department of Applied Mathematics and Statistics, Johns Hopkins University}
\affil[2]{Department of Statistics, Texas A$\&$M University}
\affil[3]{Departments of Neurology and Psychiatry, Johns Hopkins University School of Medicine}
\affil[4]{Department of Epidemiology, Johns Hopkins Bloomberg School of Public Health}
\affil[5]{\mbox{Department of Medicine, Division of Infectious Disease and Travel Medicine, Georgetown University}}
\affil[*]{Correspondence should be addressed to email: yanxun.xu@jhu.edu}
\date{}                     
\begin{document}

\maketitle

\vspace{-1cm}

\begin{abstract}
Although combination antiretroviral therapy (ART) is highly effective in suppressing viral load for people with HIV (PWH), many ART agents may exacerbate  central nervous system (CNS)-related adverse effects including depression. Therefore, understanding the effects of ART drugs on the CNS function, especially mental health, can help clinicians personalize medicine with less adverse effects for PWH and prevent them from discontinuing their ART to avoid undesirable health outcomes and increased likelihood of HIV transmission. The emergence of electronic health records offers researchers unprecedented access to HIV data including individuals' mental health records, drug prescriptions, and clinical information over time. However, modeling such data is very challenging due to high-dimensionality of the drug combination space, the individual heterogeneity, and sparseness of the observed drug combinations. We develop a Bayesian nonparametric approach to learn drug combination effect on mental health in PWH adjusting for socio-demographic, behavioral, and clinical factors. The proposed method is built upon the subset-tree kernel method that represents drug combinations in a way that synthesizes known regimen structure into a single mathematical representation. It also utilizes a distance-dependent Chinese restaurant process to cluster heterogeneous population while taking into account individuals' treatment histories. We evaluate the proposed approach through simulation studies, and apply the method to a dataset from the Women’s Interagency HIV Study, yielding interpretable and promising results. Our method has clinical utility in guiding clinicians to prescribe more informed and effective personalized treatment based on individuals' treatment histories and clinical characteristics. 


\end{abstract}

\section{Introduction}

Early initiation and adherence to antiretroviral therapy (ART) regimens optimizes health outcomes in people with HIV (PWH) and prevents further HIV transmission \citep{bangsberg2001non, de2002impact, yun2005antidepressant, saag2018antiretroviral}. 
However, viral rebound is possible due to the high viral evolutionary dynamics and the occurrence of drug-resistant mutations, ultimately resulting in treatment failure. 
ART agents fall into several classes including nucleotide reverse transcriptase inhibitor (NRTI), non-nucleotide reverse transcriptase inhibitor (NNRTI), protease inhibitor (PI), integrase inhibitor (INSTI), and entry inhibitor (EI). 
Different drug classes target HIV via different mechanisms. 
While each individual ART drug is susceptible to certain resistant mutations, a combination of drugs from different drug classes can successfully suppress the virus.
Therefore, modern ART regimens typically combine three or more drugs of different classes.

Despite the remarkable success of effective ART reducing disease-related morbidity and mortality \citep{trickey2017survival}, numerous observational studies have reported ART-related adverse effects on central nervous system (CNS) function including depression, anxiety, sleep disturbances, suicidal ideation, developmental disorders, and neurological toxicities \citep{rekha2018case, zash2018neural}. For example, significant neuropsychiatric side effects such as nightmares, hallucinations, and depression \citep{gaida2016review}, have been reported for EFV (efavirenz), an NNRTI. Peripheral neuropathy has been reported for D4T (stavudine), an NRTI, especially when used in combination with other NRTIs \citep{arenas2016peripheral, saag2018antiretroviral}.  These side effects may result in ART discontinuation with downstream consequences such as difficulties in work performance and functioning, HIV disease progression, and increased likelihood of HIV transmission \citep{fazeli2015physical, watkins2015cognitive}. Since ART is recommended for PWH indefinitely, it is critical to understand and quantify drug effects, especially the drug combination effect, on CNS function to facilitate the design and effectiveness of ART regimens.

In this paper, we focus on ART-related effect on depressive symptoms.
Depression is one of the leading mental health comorbidities in PWH, affecting from  20\% to 60\% of those with the virus \citep{bengtson2016disparities}. 
Depression is associated with numerous adverse consequences including poor ART adherence \citep{chattopadhyay2017cognitive}, rapid disease progression \citep{ironson2017depression}, and increased risk-taking behaviors \citep{brickman2017association}. 
The high prevalence and the harmful effect of depression among PWH highlights the need for effective clinical management and adequate treatment for depression.
To date, few studies are dedicated to investigating the effects of ART regimens on depression, many of which present inconsistent findings. 
For instance, \cite{pearson2009one} and \cite{mollan2014association} reported increased ART-related depression, whereas \cite{okeke2013aids} and \cite{doi:10.1080/09540120412331319714} reported the opposite.
One possible explanation is that the effects of ART regimens are heterogeneous and may be confounded by numerous factors such as socioeconomic status, behavioral factors, and clinical performance.
ART may alleviate depressive symptoms for some individuals through viral suppression and physical health improvement. 
For others, ART-related neurotoxicities may aggravate depression and lead to treatment failure. 
Therefore, investigating the heterogeneous effects of ART on depression among PWH while accounting for major confounders can help identify individual factors that drive depression with ART exposure, thereby facilitating precision medicine for PWH.

\begin{figure}[!htb]
    \centering
    \begin{tabular}{cc}
    \includegraphics[width=.5\textwidth]{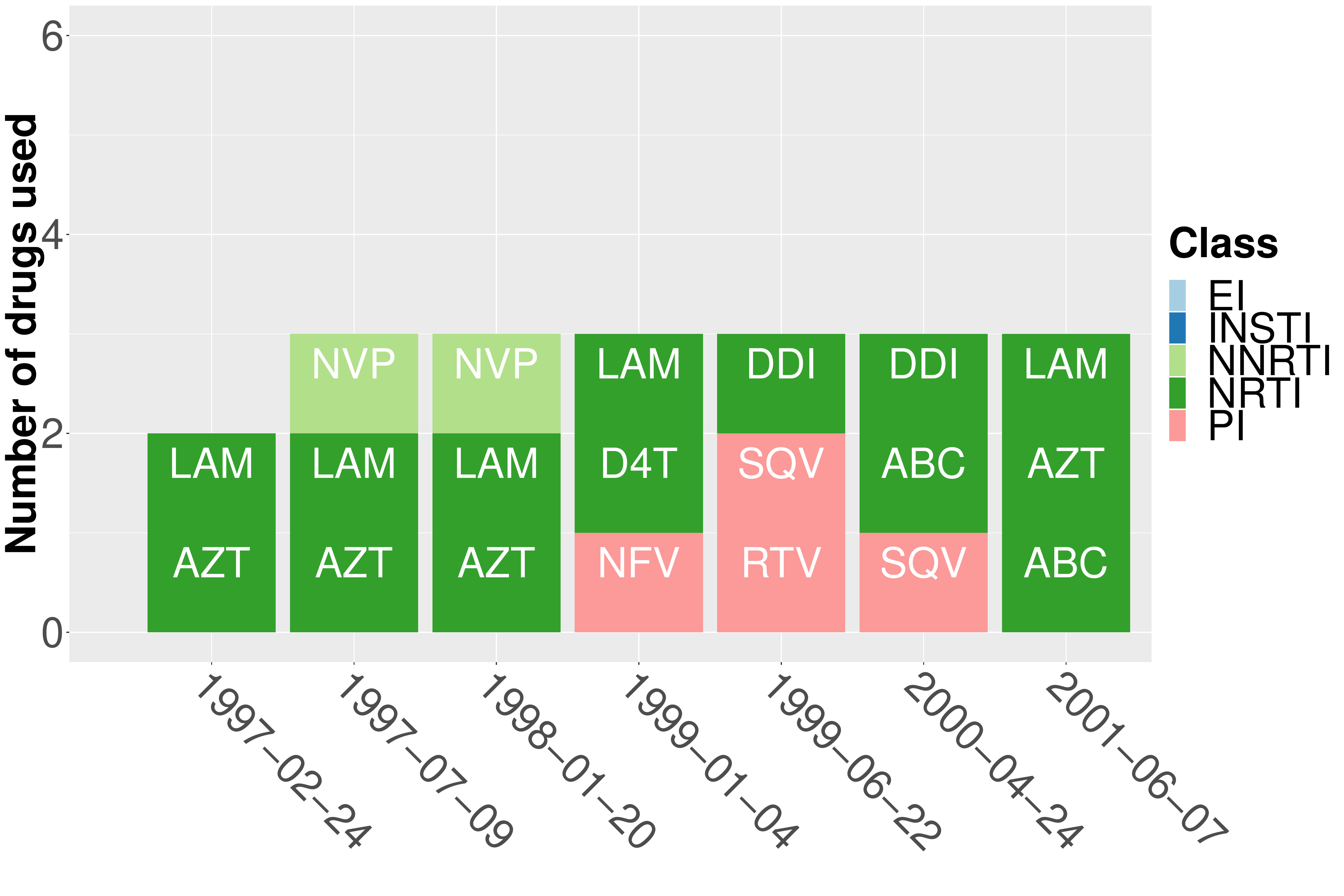} & 
    \includegraphics[width=.5\textwidth]{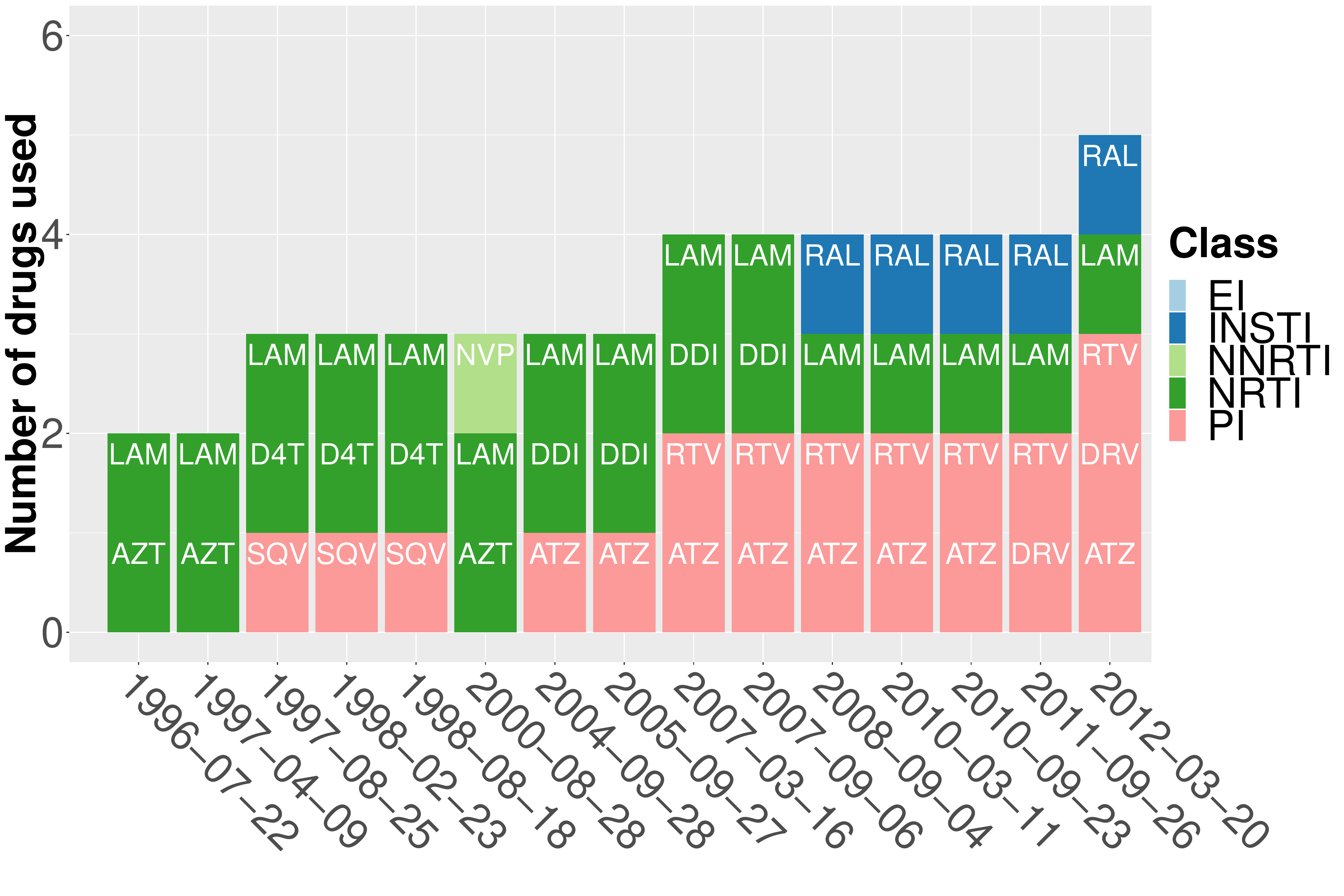} \\
	(a) ART use for individual $\#$1 & (b) ART use for individual $\#$2 \\ 
	\includegraphics[width=.5\textwidth]{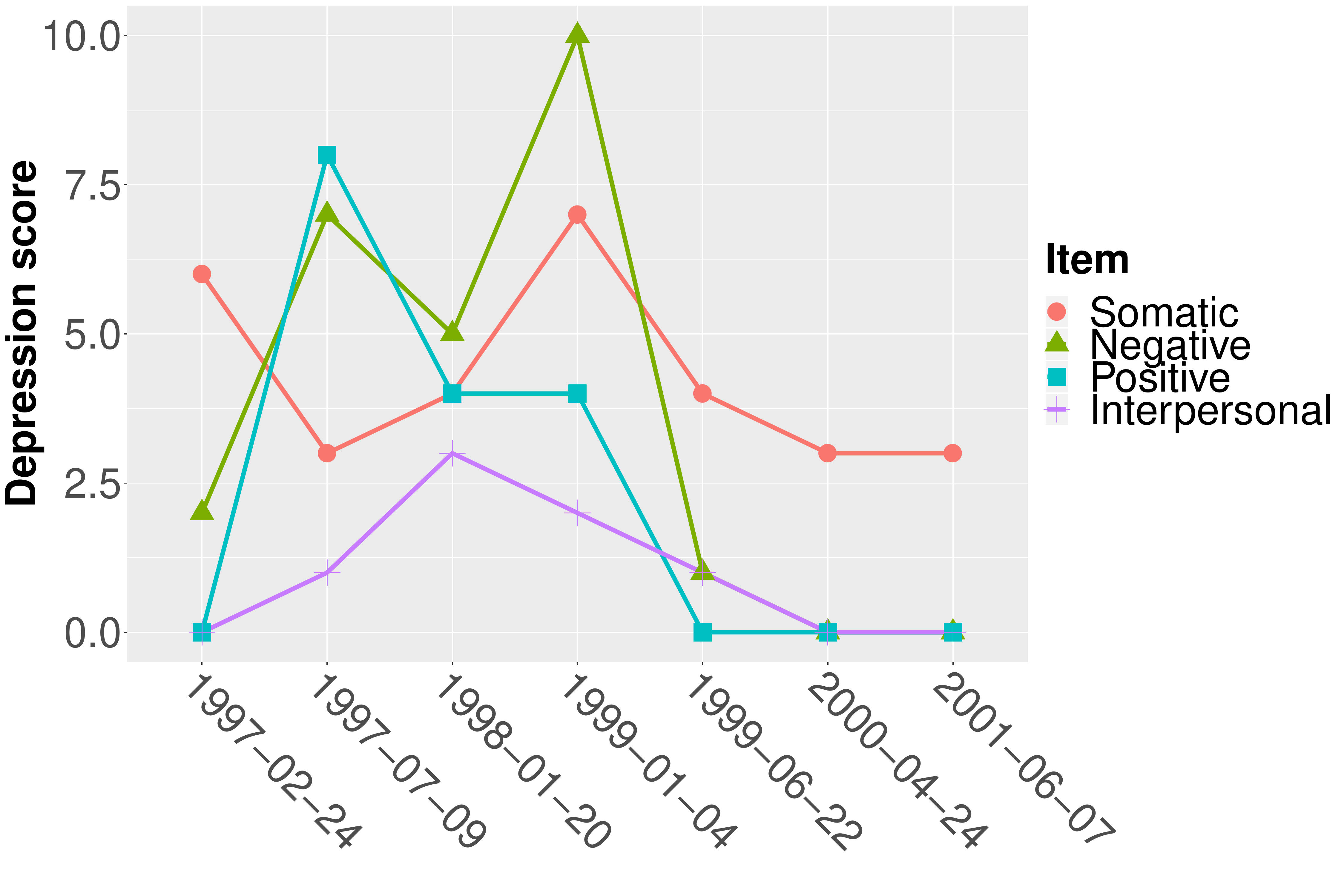} & 
	\includegraphics[width=.5\textwidth]{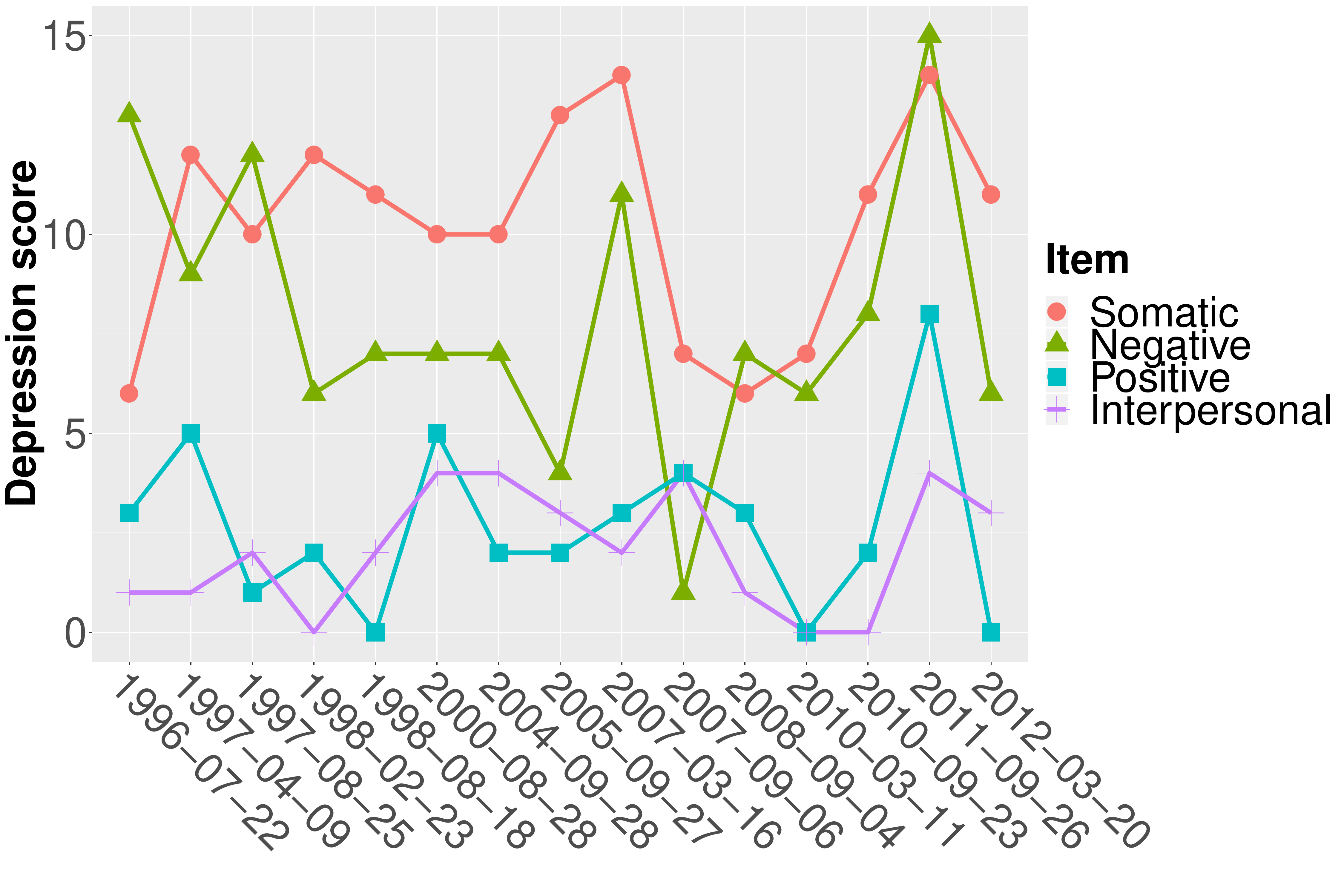} \\
    (c) Depression score for individual $\#$1 & (d) Depression score for individual $\#$2 \\
    \end{tabular}
    \caption{ART medication data and depression scores of two individuals versus their clinical visits denoted by calendar dates.}
    \label{fig:data}
\end{figure}

Large-scale HIV datasets, such as the Women’s Interagency HIV Study (WIHS), provide an opportunity and challenge to study the effects of ART regimens on depressive symptoms.
The WIHS is a prospective, observational, multicenter study which includes women with HIV and women at-risk for HIV infection in the United States \citep{bacon2005women}.
Sociodemographics, medication use, clinical diagnoses, and laboratory test results are collected longitudinally with the goal of investigating the impact of HIV infection on multimorbidity. 
For example, Figure \ref{fig:data}(a, b) presents two individuals' ART medication data versus their clinic visits denoted by calendar dates. They were followed for different time periods with distinct visit dates and drug uses. 
Their corresponding four depression scores were also recorded at each visit measuring somatic symptoms, negative affect, lack of positive affect, and interpersonal symptoms. The depression scores were summarized from a self-report questionnaire using the Center for Epidemiological Studies Depression Scale where a higher score reflects worse symptoms (CES-D, \citealt{radloff1977ces, carleton2013center}), as shown in Figure \ref{fig:data}(c, d).

The complexity of longitudinal observations, a heterogeneous population, and dynamic and mixed ART assignments present four analytical and modeling challenges.
The first challenge is high-dimensionality. With more than 20 ART drugs on the market, there are nearly a half million possible drug combinations, making the estimation of drug combination effect a high-dimensional problem.
The second issue is unbalancedness. Some ART combinations are frequent whereas others are rare. For example, D4T+LAM+NFV (two NRTIs + one PI) was recorded 993 times in the WIHS, while a similar ART regimen D4T+LAM+ATZ was only recorded 12 times.
The third issue is sparseness. Only a tiny portion (hundreds of drug combinations) of the high-dimensional ART regimen space were observed in the WIHS, while the inference on the entire space is desired.
Finally, there is the issue of non-stationarity. Since PWH are given different ART regimens during the course of treatment, the effect of the current regimen on depression is likely  dependent on prior ART regimen use. 

Statistical learning methods, such as logistic regression, tree-based models, and neural networks have been used to study the effect of ART regimens on survivals and to predict virological responses \citep{altmann2007improved, larder2007development, altmann2009predicting, caniglia2017comparison}. 
However, the representations of ART regimens in these models were simplistic, either using a binary variable to indicate whether an individual is on ART, or lumping ART regimens together into a few coarse types.
For example, \cite{lundgren2002clinically} dichotomizes ART regimens into those with or without a PI.
\cite{bogojeska2010dealing} proposed to predict binary virological responses to ART regimens by fitting a separate logistic regression model for each regimen and borrowing information from similar regimens. 
The similarity between regimens was defined by a linear additive function.
Although computationally efficient, the linear approach treats drugs as exchangeable and does not account for drug classes. Moreover, it cannot be easily extended for modeling ordinal or continuous outcomes, adjusting for covariates, and considering treatment histories in a longitudinal setup. 

To address the aforementioned challenges, we develop a novel Bayesian nonparametric model using and extending subset-tree kernels \citep{collins2002convolution} and the distance-dependent Chinese restaurant process \citep{blei2011distance, dahl2017random} to estimate the effects of ART regimens on depressive symptoms after adjusting for relevant socio-demographic, behavioral, and clinical factors. 
The subset-tree kernel method represents drug combinations in a way that synthesizes known regimen structure and the corresponding drugs into a single mathematical representation to induce an appropriate similarity among different ART regimens. 
This formulation enables us to efficiently borrow information across ART regimens and develop inferences and predictions on unobserved ART regimens. Furthermore, we extend the subset-tree kernel among ART regimens to sequences of ART regimens in a longitudinal setup, and use the distance-dependent Chinese restaurant process as a prior to capture heterogeneity among individuals by considering individuals’ treatment histories. The R code implementing our model can be found at 
\url{https://drive.google.com/open?id=1FB8o0cHx0lVq-PdEGZciVoknB8nCUXCI}. 

The rest of the paper proceeds as follows. 
In Section \ref{sec:model}, we present the proposed Bayesian nonparametric model along with the posterior inference. We evaluate the performance of the proposed approach through simulation studies and sensitivity analyses in Section \ref{sec:simu}. 
In Section \ref{sec:real}, we apply the proposed model to a large-scale HIV clinical dataset to study the effects of ART regimens on depressive symptoms. Finally, we conclude with a discussion in Section \ref{sec:dis}.

\section{Model and Inference}
\label{sec:model}

\subsection{Probability Model}

Denote $Y_{ijq}$ to be the score of depression item $q$ for individual $i$ at visit $j$, 
where $i=1, \dots, n$, $j=1, \dots, J_i$, and $q=1, \dots, Q$.
Let $Z_{ij}$ denote the ART regimen used by individual $i$ at visit $j$.
For example, $Z_{ij}=$ D4T+LAM+NFV if individual $i$ takes a combination of drugs D4T, LAM, and NFV at visit $j$. 
Let $\bX_{ij}$ be an $S$-dimensional vector including an intercept, time-invariant covariates (e.g., race), and time-varying covariates (e.g., BMI, CD4 count) for individual $i$ at visit $j$. 
We construct a sampling model for the depression score $\bm{Y}_{ij} = (Y_{ij1},\dots,Y_{ijQ})^{\mT}$ as follows,
\begin{equation}\label{eqn:sampling_model}
    \bm{Y}_{ij} =  \bm{\beta}_{i}\bm{X}_{ij} + \bm{h}(Z_{ij}) + \bm{\omega}_{ij} + \bm{\varepsilon}_{ij},
\end{equation}
where $\bm{\beta}_{i}$ is a $Q\times S$ dimensional matrix, $\bm{h}(\cdot)$ is a $Q$-dimensional vector-valued function, $\bm{\omega}_{ij}$ is a $Q$-dimensional vector following a multivariate normal distribution $\mathcal{N}(\bm{0}, \sigma_{\varepsilon}^2 \bm{\Sigma}_{\omega})$ that models the dependency among different depression items, and $\bm{\varepsilon}_{ij} \sim \mathcal{N}(\bm{0}, \sigma_{\varepsilon}^2 \bm{I}_Q)$ is an independent normal error. 
For identifiability, we assume $\bm{\Sigma}_{\omega}$ to be a correlation matrix.  
The first term $\bm{\beta}_{i}\bm{X}_{ij}$ in \eqref{eqn:sampling_model} captures the dependence of the outcome $\bm{Y}_{ij}$ on the covariate $\bm{X}_{ij}$. 
The second term $\bm{h}(Z_{ij})$ is the key component of our model characterizing the combination effect of ART regimen $Z_{ij}$, the details of which are given below.

\underline{\textit{Combination effect $\bm{h}(\cdot)$.}} We construct $\bm{h}(\cdot)$ with two desired properties: 1) sharing-of-information - encouraging similar effects for similar ART regimens; and 2) parsimony - reducing the high dimensional ART regimen space to a manageable size.
Specifically, we first pick a number $D$ of representative ART regimens, denoted by $z_1, \dots, z_D$, which are similar to the notion of knots in splines. 
Then we model the  combination effect of ART regimen $Z_{ij}$ using a kernel smoother,
\begin{equation}\label{eqn:combined_drug_kernel}
    \bm{h}(Z_{ij}) = \frac{\sum_{d=1}^{D} \kappa(Z_{ij}, z_d)\bm{\gamma}_{id}}{\sum_{d=1}^{D} \kappa(Z_{ij}, z_d)},
\end{equation}
where $\bm{\gamma}_{id}$'s are $Q$-dimensional vectors and the kernel weights are defined by an ART regimen similarity function $\kappa(Z_{ij}, z_d)$. How to choose these representative regimens will be discussed later in the simulation studies and applications. 
The concept of similarity between different regimens has been introduced in HIV studies. 
\cite{bogojeska2010dealing} proposed a linear kernel method to compute the similarity between regimens based on the proportion of common drugs that two regimens share, $\kappa(z_d, z_{d'})=(u_{z_d}^{T} u_{z_{d'}})/\max(u_{z_d}^{T}\bm{1}, u_{z_{d'}}^{T}\bm{1})$, where $u_{z_d}$ is a binary vector indicating the drugs comprising the regimen $z_d$ and $\bm{1}$ is a vector of 1's.
Although conceptually simple, the linear kernel treats all ART drugs as exchangeable and does not account for drug classes.
For example, the regimen D4T (NRTI) + LAM (NRTI) + NFV (PI) should be more similar to D4T + LAM + ATZ (PI) than D4T + LAM + EFV (NNRTI) since NFV and ATZ belong to the same drug class PI whereas EFV belongs to another class NNRTI. 
However, the linear kernel approach gives rise to the same similarity score for these two pairs.

\begin{figure}[!htb]
    \centering
    \begin{tabular}{cc}
    \includegraphics[width=5cm,height=5cm,clip,keepaspectratio]{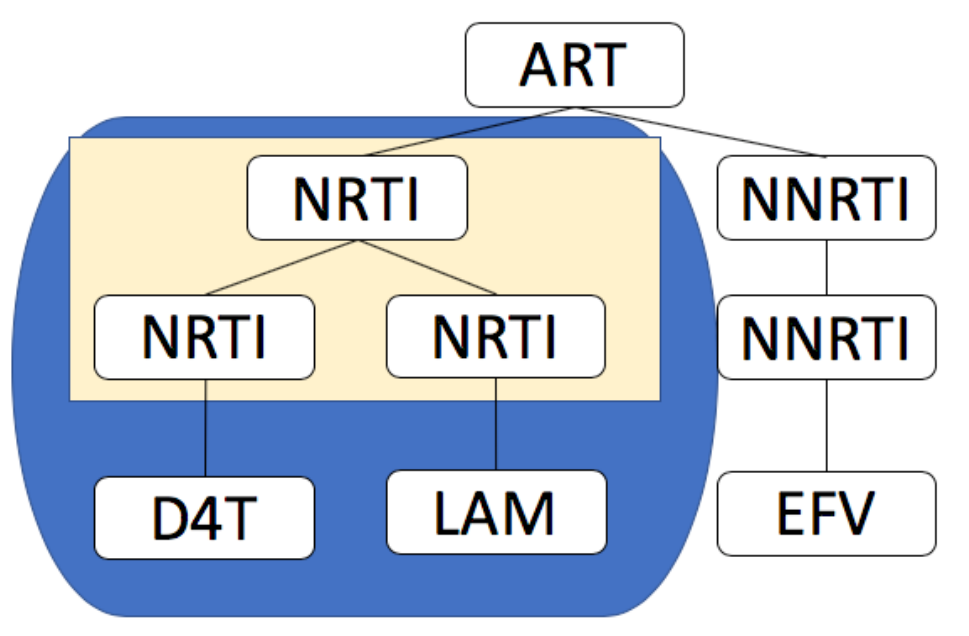} & 
    \includegraphics[width=5cm,height=5cm,clip,keepaspectratio]{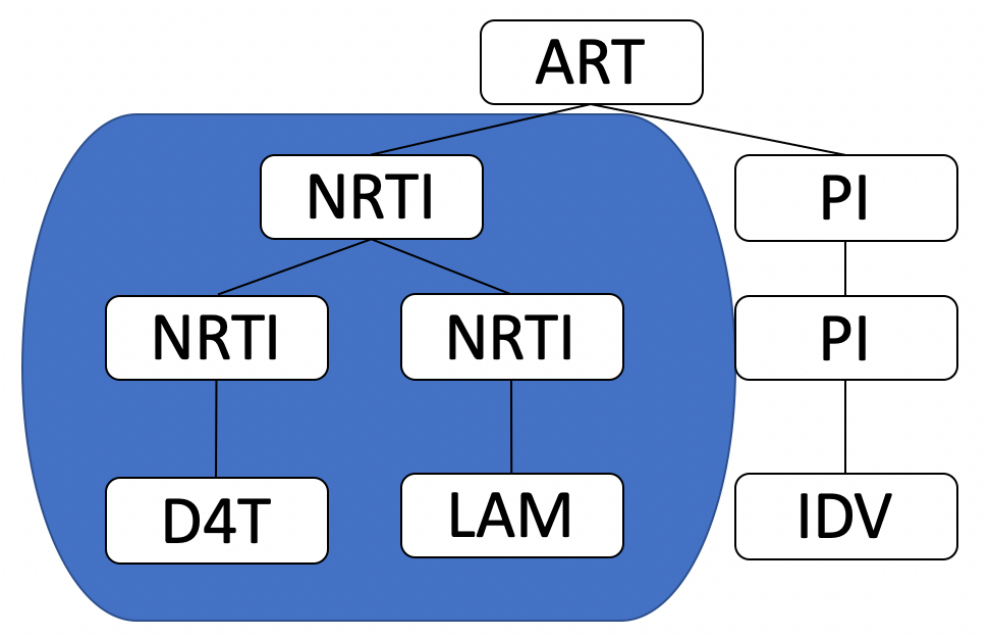} \\
	(a) Regimen A & (b) Regimen B \\ 
	\includegraphics[width=5cm,height=5cm,clip,keepaspectratio]{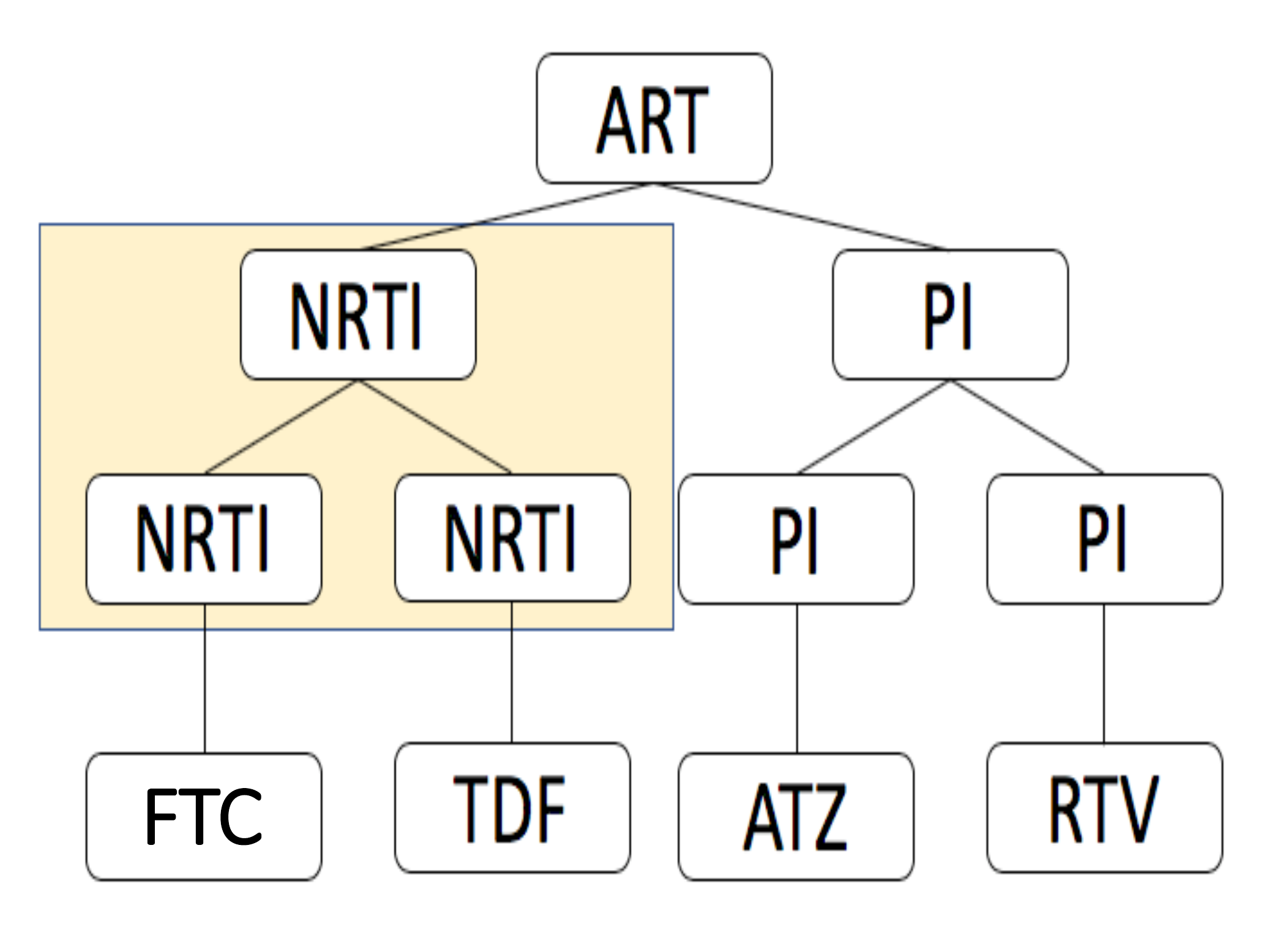} & 
	\includegraphics[width=5cm,height=5cm,clip,keepaspectratio]{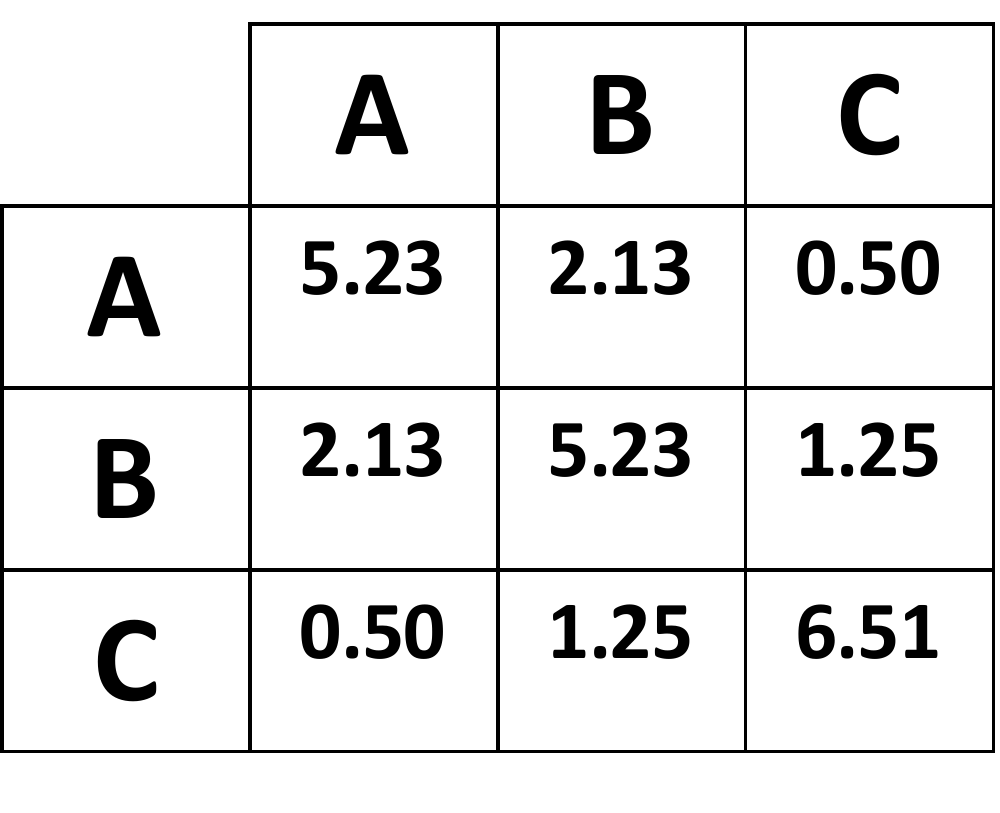} \\
    (c) Regimen C & (d) Similarity score matrix \\
    \end{tabular}
    \caption{Tree representations for ART regimens with their similarity matrix.}
    \label{fig:drug_simi}
\end{figure}

We propose to use a subset-tree (ST) kernel method, which was originally developed in natural language processing to represent sentence
structure \citep{collins2002convolution}.
We represent each ART regimen as a rooted tree which encodes the knowledge about the regimen structure such as  drug classes and the number of distinct drug classes under each regimen. 
Figure \ref{fig:drug_simi} illustrates this idea using three regimens (A, B, C) as an example.  
Regimen A contains D4T (NRTI) + LAM (NRTI) + EFV (NNRTI); regimen B contains D4T (NRTI) + LAM (NRTI) + IDV (PI); and regimen C contains FTC (NRTI) + TDF (NRTI) + ATZ (PI) + RTV (PI). 
The linear kernel used in \cite{bogojeska2010dealing} would assign 0 similarity score to regimens A and C since they share no common drugs.
But the ST kernel will be able to capture the similarity on the drug class level (highlighted by the yellow boxes).
In fact, ST kernel calculates the similarity score between regimens across all levels of the tree representation.
This feature will later be further exploited to compute the similarity between (longitudinal) sequences of regimens of possibly different lengths.

The main idea of the ST kernel is to compute the number of common substructures between two trees $T_a$ and $T_b$. 
Let $R_T$ denote the set of nodes for any tree $T$ and let $ch(r)$ denote the set of children nodes of the node $r \in R_T$ (i.e., nodes immediately below $r$).
The similarity score, $\kappa(T_a, T_b)$ between two regimen trees $T_a$ and $T_b$, is calculated by
\begin{equation}\label{eqn:st_kernel}
    \kappa(T_a, T_b) = \sum_{r_a\in R_{T_a}}\sum_{r_b\in R_{T_b}}\rho(r_a, r_b),
\end{equation}
where $\rho(r_a, r_b)$ is defined for each pair of nodes as follows.
(i) If $r_a$ and $r_b$ are terminal nodes $(ch(r_a) = ch(r_b) = \emptyset)$, then $\rho(r_a, r_b) = 0$.
(ii) If $r_a$ and $r_b$ have different sets of children nodes $(ch(r_a) \not= ch(r_b))$, then $\rho(r_a, r_b) = 0$.
(iii) If $r_a$ and $r_b$ have the same nonempty set of children nodes, then 
$\rho(r_a, r_b) = \eta \prod_{s=1}^{|ch(r_a)|}\left\{1+\rho(ch_{r_a}^{s}, ch_{r_b}^{s})\right\}$,
where $|\cdot|$ is the cardinality of a set and $ch_{r_a}^{s}$ is a child of $r_a$ for $s=1,\dots,|ch(r_a)|$. 
Here we include a hyperparameter $\eta \in (0,1]$, which is a decay factor to control the relative influence from nodes near the root to alleviate the peakiness of the ST kernel when the depth of tree fragments is considerably large \citep{beck2015learning}. 
Figure \ref{fig:drug_simi}(d) presents the similarity score matrix among A, B, and C when $\eta = 0.5$. 
Note that the self-similarity of regimen C is higher than those of regimens A or B because regimen C consists of more drugs and therefore has a higher similarity score due to the additive definition.

To estimate $\bm{\gamma}_{id}$'s in \eqref{eqn:combined_drug_kernel}, another challenge arises from the potential high-dimensionality and multicollinearity of the kernel weights calculated from the similarity function. 
Following the idea from principal component regression \citep{kendall1965course}, we consider a principal component analysis on the design matrix that consists of kernel weights. Specifically, let $\bm{H}_{ij}$ be a $D$-dimensional vector whose $d$-th element is the kernel weight $\kappa(Z_{ij}, z_d)/\sum_{d=1}^{D}\kappa(Z_{ij}, z_d)$, and let $\bm{H} = (\bm{H}_{11}^{\mT}, \dots, \bm{H}_{1J_1}^{\mT}, \dots, \bm{H}_{n1}^{\mT}, \dots, \bm{H}_{nJ_n}^{\mT})^{\mT}$ be the $N \times D$ design matrix ($N=\sum_{i=1}^{n}J_i$) for the kernel regression in \eqref{eqn:combined_drug_kernel}. We perform the principal component analysis on $\bm{H}$ and retain the first $D^\star$ principal components that explain at least 99.9\% of the total variance. The resulting $N\times D^{\star}$ matrix is denoted by $\tilde{\bm{H}}$.
Then the combination effect $\bm{h}(\cdot)$ can be approximated by $\bm{\gamma}_{i}^\star\tilde{\bm{H}}_{ij}$, where $\bm{\gamma}_{i}^\star$ is the $Q\times D^\star$ matrix that needs to be estimated.

\subsection{Priors}

To capture heterogeneity among individuals and  individual treatment histories, we use the distance-dependent Chinese restaurant process (ddCRP, \citealt{blei2011distance, dahl2017random}) to induce clustering of individuals depending on the similarity between their treatment histories. 
Let $\bm{\theta}_i = \{\bm{\beta}_{i}, \bm{\gamma}^\star_{i}\}$. We assume that $\bm{\theta}_i \sim \text{ddCRP}(m_0, s, G_0)$ with a mass parameter $m_0$, a base measure $G_0$, and a similarity function $s(\cdot,\cdot)$.  

We give a brief introduction to the ddCRP below. 
Due to the discrete nature of ddCRP, $\{\bm{\theta}_i\}_{i=1}^n$ are likely to have ties. Let $\{\btth_k\}_{k=1}^{r_n}$ denote the $r_n$ unique values of $\{\bm{\theta}_i\}_{i=1}^n$ where $\btth_k=\{\tilde{\bm{\beta}}_{k}, \tilde{\bm{\gamma}}_{k}^\star\}$. 
Let $\pi_n=\{S_1,\dots, S_r\}$ denote a partition of $[n]=\{1,2,\dots,n\}$ such that $\cup_{k=1}^{r_n}S_k=[n]$. The ties among $\bm{\theta}_i$'s naturally give rise to a partition, i.e., $\bm{\theta}_i=\btth_{k}$ if individual $i$ belongs to cluster $k$, $i\in S_k$.
Following \cite{dahl2017random}, $\bm{\theta}_i \sim \text{ddCRP}(m_0, s, G_0)$ can be written as
\begin{equation}\label{eqn:ddcrp}
    \bm{\theta}_i = \sum_{k=1}^{r_n}\btth_{k}I(i\in S_k), \quad \btth_{k}\sim G_0, \quad \pi_n \sim p(\pi_n).
\end{equation}

Denote $\bm{\sigma}=(\sigma_{1},\dots,\sigma_{n})$ a permutation of $[n]$, $\pi(\sigma_{1},\dots,\sigma_{t-1})$ a  partition of $\{\sigma_{1},\dots,\sigma_{t-1}\}$, and $r_{t-1}$ the number of subsets in $\pi(\sigma_{1},\dots,\sigma_{t-1})$, $t\leq n$. 
The probability mass function of the partition $\pi_n$ is defined as the product of increasing conditional probabilities \citep{dahl2017random},
\begin{equation}\label{eqn:partition_pmf_1}
    p(\pi_n\mid m_0, s, \bm{\sigma}) = \prod_{t=1}^{n}p_t(m_0, s, \pi(\sigma_{1},\dots,\sigma_{t-1})),
\end{equation}
where 
\begin{align}\label{eqn:partition_pmf_2}
    p_t(m_0, s, \pi(\sigma_{1},\dots,\sigma_{t-1}))  
    & = \text{Pr}(\sigma_t\in S\mid m_0, s, \pi(\sigma_{1},\dots,\sigma_{t-1})) \nonumber \\
    & = \begin{cases}
        &  \frac{t-1}{m_0+t-1}\frac{\sum_{\sigma_s\in S}s(\sigma_t,\sigma_s)}{\sum_{s=1}^{t-1}s(\sigma_t,\sigma_s)}, 
        \text{ for } S\in \pi(\sigma_{1},\dots,\sigma_{t-1}) \\
        & \frac{m_0}{m_0+t-1}, \text{~~~~~~~~~~~~~~~~ for } S \text{ being a new subset,}
    \end{cases}
\end{align}
and $p_t(m_0, s, \pi(\sigma_{1},\dots,\sigma_{t-1}))=1$ for $t=1$ by convention.

\begin{figure}[!htb]
    \centering
    \begin{tabular}{c}
    \includegraphics[width=10cm,height=5cm,clip,keepaspectratio]{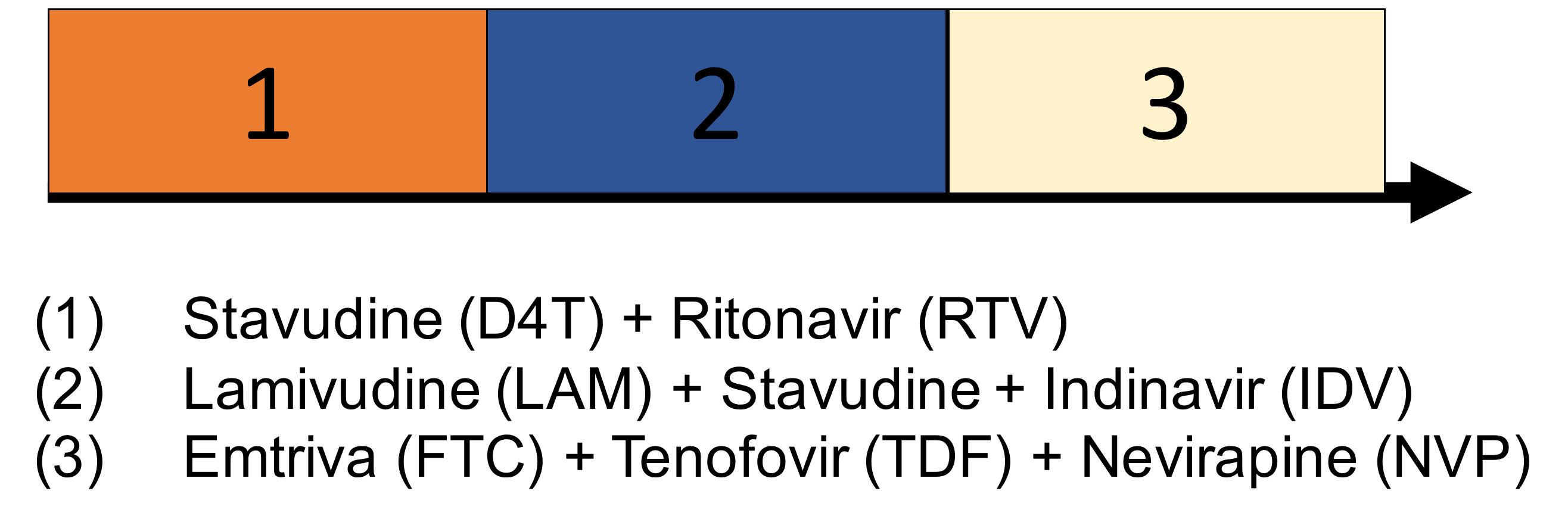} \\
    \includegraphics[width=10cm,height=5cm,clip,keepaspectratio]{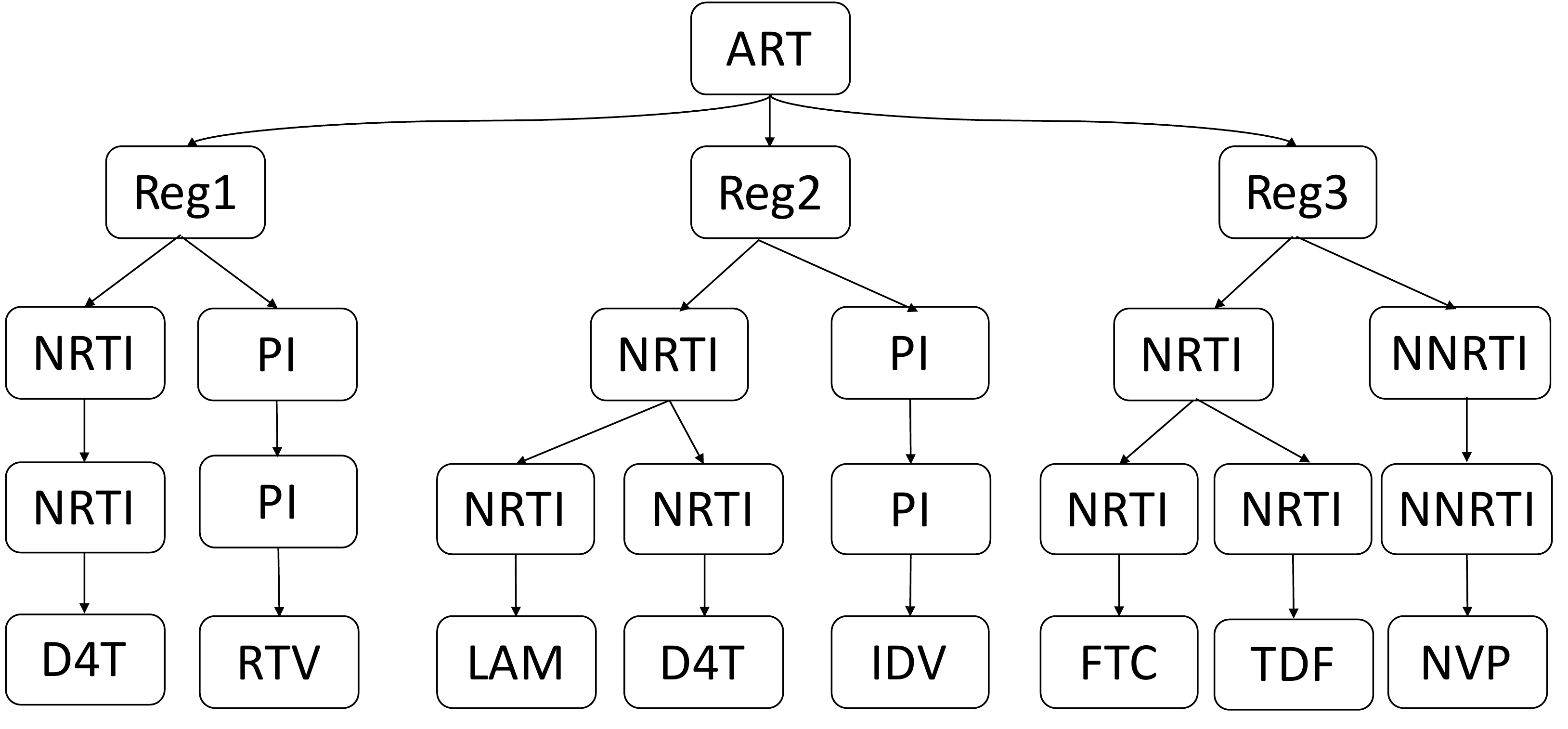} \\
    \end{tabular}
    \caption{Tree representation (bottom) for a sequence of ART regimens (top).}
    \label{fig:seq_drug}
\end{figure}

\underline{\it Similarity function $s(\cdot,\cdot)$.} 
The similarity function $s(\cdot,\cdot)$ depends on individuals' treatment histories. 
Let $T_i$ and $T_{i'}$ denote two sequences of treatment regimens for individuals $i$ and $i'$, respectively. 
The proposed ddCRP prior assumes that the prior probability that they belong to the same cluster is proportional to 
the similarity score $s(i,i')=\kappa(T_i,T_{i'})$. 
Therefore, individuals with similar ART regimen histories are \textit{a priori} more likely to be clustered together.
To measure the similarity between two regimen sequences, we extend the ST kernel in $\eqref{eqn:st_kernel}$ by combining multiple regimen trees into a single tree under the common root ``ART.''
Figure \ref{fig:seq_drug} shows an example of the tree structure for one ART regimen sequence with three distinct ART regimens.
Then we apply the ST kernel in $\eqref{eqn:st_kernel}$ to calculate the similarity score between regimen sequences in the same fashion as before. 

We complete the model by assigning hyperpriors. We use a conjugate gamma prior $\mathrm{Gamma}(c_0, d_0)$ on $m_0$, a conjugate inverse-gamma prior $\text{Inverse-Gamma}(g_1, g_2)$ on $\sigma_{\varepsilon}^2$, and a uniform distribution on the permutation, i.e., $p(\bm{\sigma})=1/n!$ for all $\bm{\sigma}$, for ease of posterior computation. 
In addition, we use a conjugate normal prior as the base measure $\btth_k=\{\tilde{\bm{\beta}}_{k}, \tilde{\bm{\gamma}}_{k}^\star\}\sim G_0$. Specifically, let $\btbeta_{kq}$ and $\btgamma_{kq}^\star$ be the $q$-th row of $\btbeta_k$ and $\btgamma_k^\star$, respectively. We assume that $\btbeta_{kq} \sim \mathcal{N}(\bm{e}_q,\bm{B}_q)$ and $\btgamma_{kq}^\star \sim \mathcal{N}(\bm{f}_q,\bm{\Lambda}_q)$ for $k=1,2,\dots,r_n$, and $q=1,2,\dots,Q$, where $\bm{e}_q \sim \mathcal{N}(\bm{0},\bm{E}_0)$, $\bm{B}_q \sim \text{Inverse-Wishart}(b_0, \bm{B}_0^{-1})$, $\bm{f}_q \sim \mathcal{N}(\bm{0},\bm{F}_0)$, and $\bm{\Lambda}_q \sim \text{Inverse-Wishart}(\lambda_0, \bm{\Lambda}_0^{-1})$. 
For the correlation matrix $\bm{\Sigma_\omega}$, we assume that $p(\bm{\Sigma_\omega}) \propto \det(\bm{\Sigma_\omega})$ following 
\cite{lewandowski2009generating}, where $\text{det}(\cdot)$ denotes the determinant of a matrix.

\subsection{Posterior Inference}

We carry out posterior inference with the Markov chain Monte Carlo (MCMC) algorithm. 
The posterior sampling for $\bm{\omega}_{ij}$'s, $\bm{\Sigma}_{\omega}$, and $\sigma_{\varepsilon}^2$ is straightforward through standard Gibbs sampler and Metropolis-Hastings sampler. To draw posterior samples for the parameters related to the ddCRP prior,  the key step is to compute  the full conditional distribution of the partition based on the probability mass function in $\eqref{eqn:partition_pmf_1}$ and $\eqref{eqn:partition_pmf_2}$.
Suppose at the current state, the partition is $\pi_n=\{S_1,\dots,S_{r_n}\}$ and let $S_k^{-}$, $k=1,2,\dots,r_n$, denote these subsets without individual $i$. 
Let $\pi_n^{i\to k}$ be the partition obtained by moving $i$ from its current subset to the subset $S_k^{-}$. 
Here we let $k=0$ denote the index of a new empty subset $S_0^{-}$, and let $\pi_n^{i\to 0}$ denote the partition after moving $i$ to a new subset. 
Then the full conditional distribution for the allocation of individual $i$ is given by,
\begin{equation*}\label{eqn:update_pi_n}
    p(i \in S_k^{-}\mid \cdot) \propto p(\pi_n^{i\to k} \mid m_0, s, \bm{\sigma})\prod_{j=1}^{J_i}p(\bm{Y}_{ij} \mid\btth_k),
\end{equation*}
for $k=0,1,\dots,r_n$, where the new parameters $\btth_0$ are drawn from the base measure  $G_0$. 
Note that $p(\pi_n^{i\to k} \mid m_0, s, \bm{\sigma})$ is calculated by evaluating $\eqref{eqn:partition_pmf_1}$ and $\eqref{eqn:partition_pmf_2}$ at the partition $\pi_n^{i\to k}$.
More details of the MCMC can be found in the Supplementary Material Section A.

\section{Simulation Study}
\label{sec:simu}

In this section, we conducted simulation studies to evaluate the performance of the proposed model by comparing the posterior inference to the simulation truth. 
To demonstrate the advantages of using the ddCRP prior for taking into account individuals' heterogeneity and treatment histories, and the ST kernel for inducing an appropriate ART regimen similarity, we compared the proposed model to two alternative methods. The first alternative replaces the ddCRP prior on $\bth_i=(\tilde{\bm{\beta}}_{i}, \tilde{\bm{\gamma}}_{i}^\star)$ with independent conjugate multivariate normal priors on $\tilde{\bm{\beta}}_{i}$'s and  $\tilde{\bm{\gamma}}_{i}^\star$'s that do not take into account individuals' heterogeneity and treatment histories, and replaces the ST kernel with a linear kernel \citep{bogojeska2010dealing} based on the proportion of common drugs that two regimens share, $\kappa(z_d, z_{d'})=(u_{z_d}^{T} u_{z_{d'}})/\max(u_{z_d}^{T}\bm{1}, u_{z_{d'}}^{T}\bm{1})$, where $u_{z_d}$ is a binary vector indicating the drugs comprising the regimen $z_d$ and $\bm{1}$ is a vector of 1's. The linear kernel only considers the number of same drugs in each pair of regimens, but ignores the drug class information. We call this method Normal+Linear. 
The second alternative, called DP+Linear, replaces the ddCRP prior on $\bth_i$ with a Dirichlet process prior that does not account for individuals' treatment histories and replaces the ST kernel with the linear kernel as in the first alternative method. Furthermore, the robustness of the decay factor $\eta$ in the ST kernel was demonstrated by sensitivity analyses.

\subsection{Simulation setup}

\begin{figure}[!htb]
    \centering
    \begin{tabular}{ll}
    \includegraphics[width=.5\textwidth]{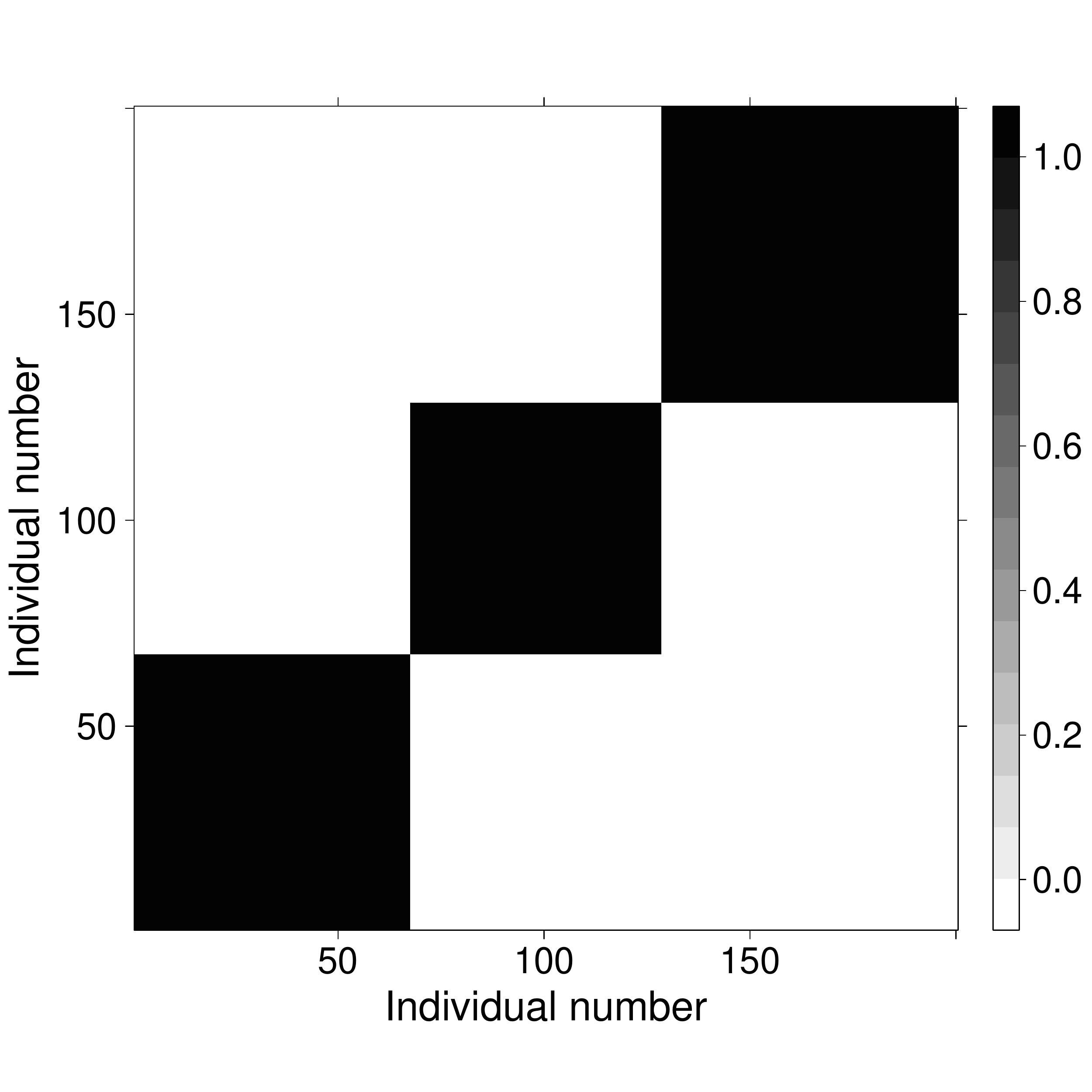} &
    \includegraphics[width=.5\textwidth]{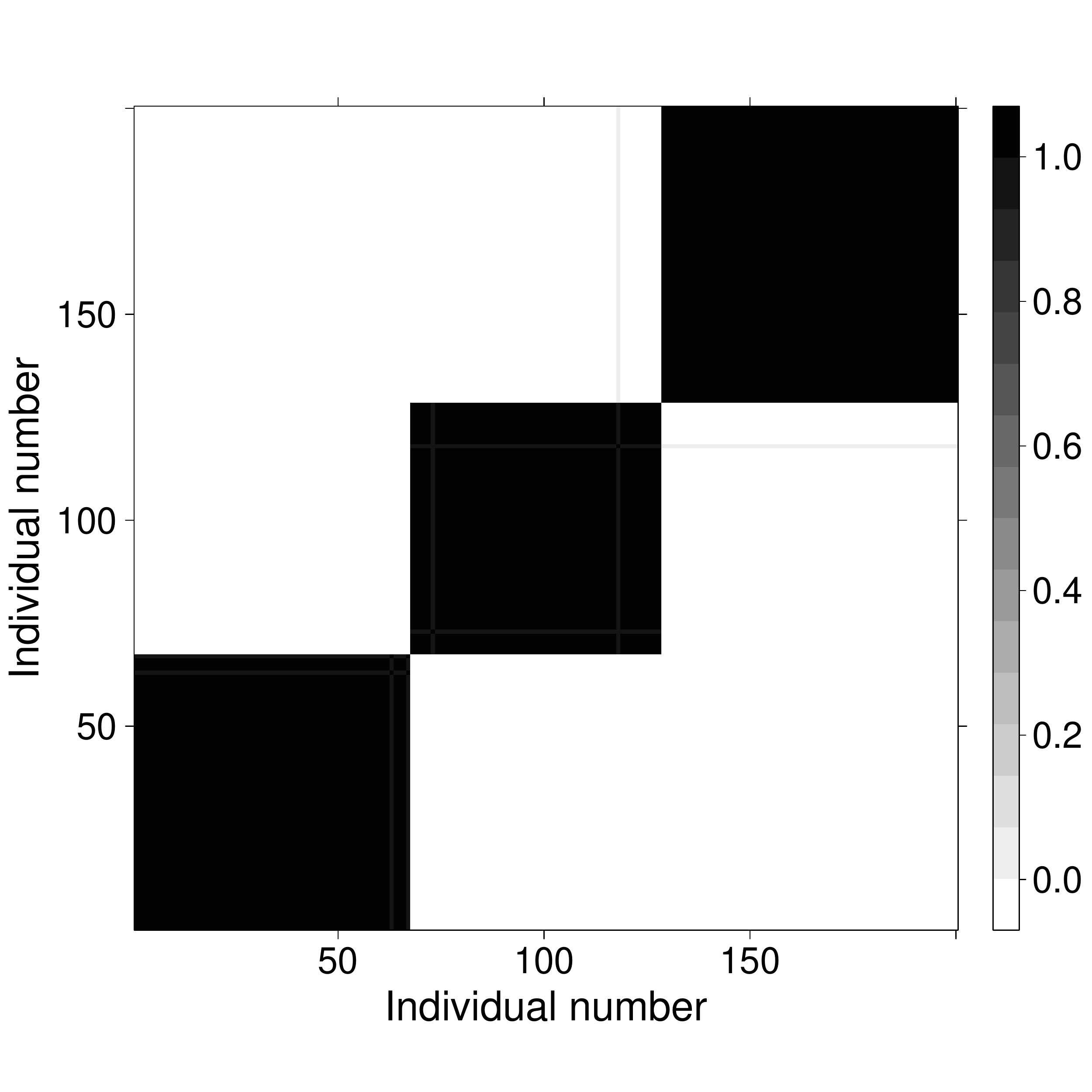} \\
    (a) Simulated true clustering scheme & 
    (b) Posterior probabilities of individual co-clustering
    \end{tabular}
    \caption{Simulated true clustering scheme and posterior probabilities of individual co-clustering averaged over 100 repeated simulations.}
    \label{fig:co_clustering}
\end{figure}

Assume that there were $n=200$ individuals with $Q=3$ depression items and $S=3$ covariates with one intercept, one time-invariant covariate, and one time-varying covariate, i.e., $\bX_{ij}=(1, x_{i0}, x_{ij})^{\mT}$, where $x_{i0}$'s and $x_{ij}$'s were generated from independent standard normal distributions, $i=1, \dots, n$, $j=1, \dots, J_i$.  
Individuals' treatment histories were randomly sampled from the WIHS dataset without replacement, resulting in the number of visits per individual to range from 2 to 38. 
We set the simulated true decay factor $\eta^o=0.5$, and then computed the similarity scores among individuals' treatment histories. 
Based on the similarity scores among different individuals, we randomly generated one realization from the ddCRP prior, yielding the simulated true number of clusters to be $r^o_n=3$ and the number of individuals in each clusters is 67, 61, and 72, respectively. 
Figure \ref{fig:co_clustering}(a) presents the simulated true clustering scheme. Conditional on the clustering memberships, we generated the simulated true $\{\btbeta_{kq}\}_{k=1}^{r^o_n},_{q=1}^{Q}$ from a standard multivariate normal distribution, the values of which are shown in Supplementary Table S1. 

We selected representative drug regimens $z_1,\dots,z_D$ if a regimen $z_d$ has been used in more than 10 visits among all the 200 individuals, yielding $D=56$.
We generated the simulated true $\{\btgamma_{kq}\}_{k=1}^{r^o_n},_{q=1}^{Q}$ in the kernel regression from a standard multivariate normal distribution, and computed the kernel weight matrix $\bm{H}$ by applying the ST kernel in \eqref{eqn:st_kernel} to individuals' treatment regimens and the selected representative drug regimens.
We performed the principal component analysis on the design matrix $\bm{H}$, and chose the first $D^{\star}=39$ principal components for $\tilde{\bm{H}}$ that explains 99.9\% variation of the original matrix. 
We set the non-diagonal elements of the correlation matrix $\bm{\Sigma}_{\omega}$ to be 
$(\sigma_{12}, \sigma_{13}, \sigma_{23}) = (0.25, 0.5, 0.75)$, and $\sigma_{\varepsilon}^2=1$.
Lastly, we generated the depression scores $\bm{Y}_{ij}$ from $\eqref{eqn:sampling_model}$. 

We applied the proposed model to the simulated dataset with 100 repeated simulations. The hyperparameters were set to be $c_0=1$, $d_0=1$, $g_1=1$, $g_2=1$, $\bm{E}_0=100\bm{I}_{S}$, $b_0=S+1$, $\bm{B}_0^{-1}=100\bm{I}_{S}$, $\bm{F}_0=100\bm{I}_{D^{\star}}$, $\lambda_0=D^{\star}+1$, and $\bm{\Lambda}_0^{-1}=100\bm{I}_{D^{\star}}$. For each analysis, we ran 10,000 MCMC iterations with an initial burn-in of 5,000 iterations and a thinning factor of 10. Convergence diagnostic assessed using R package \pkg{coda}, including autocorrelation plots and trace plots (Figures S2 and S3 in the Supplementary Material) of the post-burn-in MCMC samples for some randomly selected parameters, showed no issues of non-convergence.

\subsection{Simulation results}

We first report on the performance in terms of recovering the individual clustering. 
Our model successfully identified $\hat{r}_n = 3$ as it only overestimated the true number of clusters by 1 in 2.24\% of the post-burn-in MCMC posterior samples among all the 100 repeated simulations.
We further calculated the posterior probabilities of individual co-clustering based on the empirical proportions of individuals being clustered in the same cluster over the post-burn-in MCMC samples. The co-clustering probability matrix averaged over 100 repeated simulations is shown in Figure \ref{fig:co_clustering}(b), indicating that the proposed method assigns individuals to their simulated true clusters with high probabilities. 
 
\begin{figure}[!htb]
    \centering
    \begin{tabular}{cc}
    \includegraphics[width=.5\textwidth]{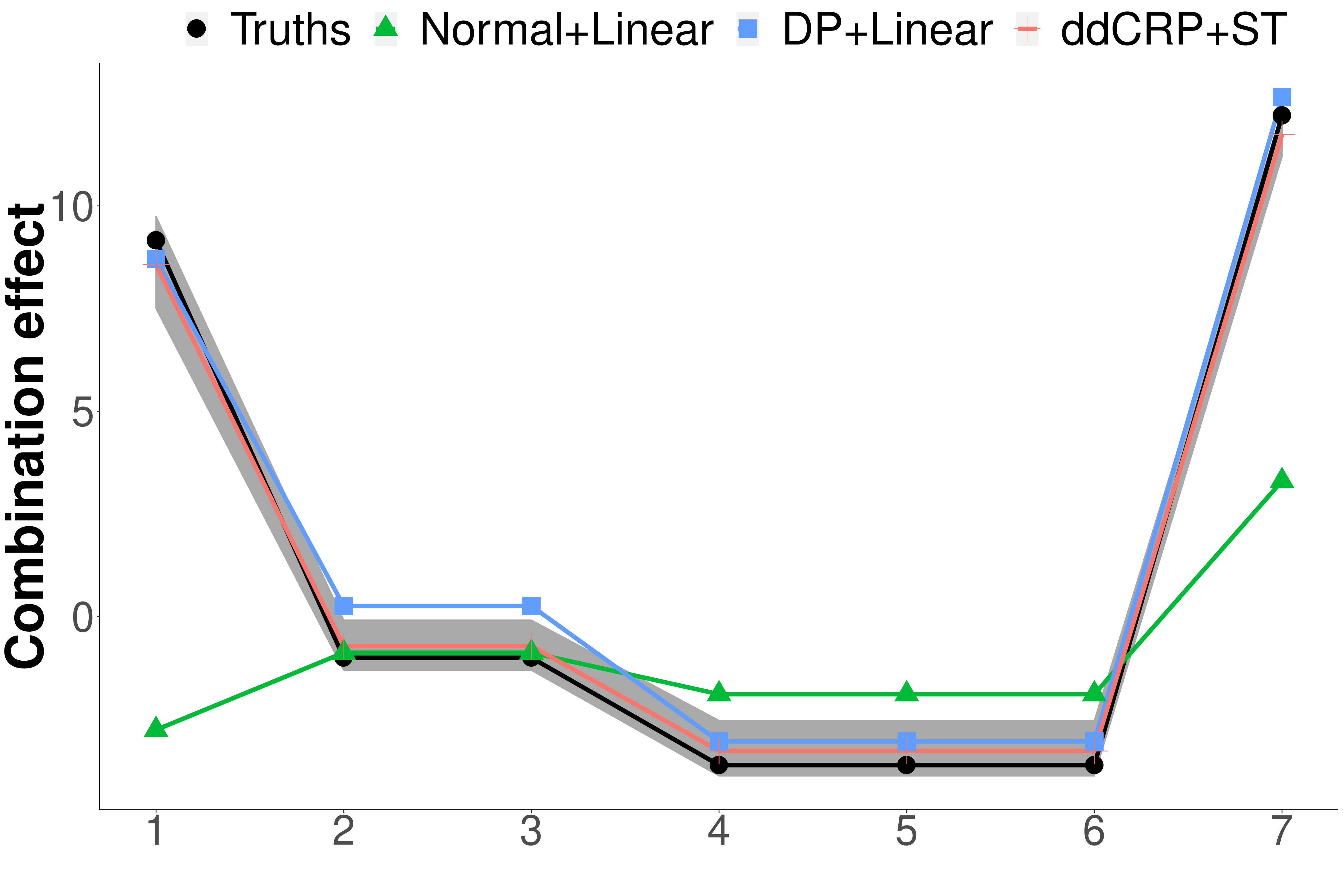} & 
    \includegraphics[width=.5\textwidth]{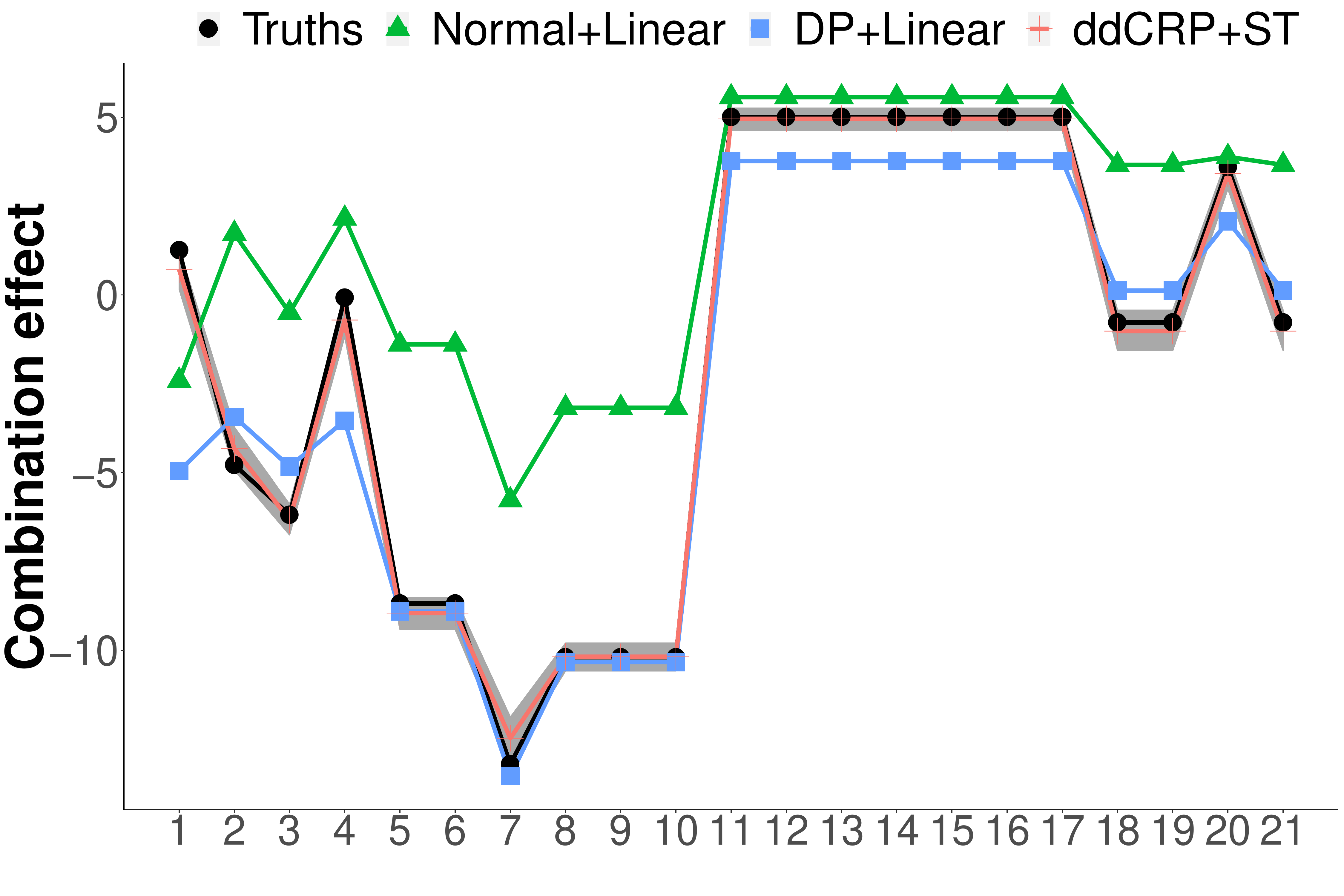} \\
	(a) Combination effects for individual $\#$1 & (b) Combination effects for individual $\#$2 \\ 
    \end{tabular}
    \caption{Combination effects for two randomly selected individuals from one randomly selected simulated dataset. The horizontal axis is the index of visit, and the vertical axis is the combination effect. The black lines represent the simulated truths of combination effects, the green lines represent the estimations under the Normal+Linear method, the blue lines represent the estimations under the DP+Linear method, and the red lines represent the estimations under the proposed method
 (ddCRP+ST). The shaded area represents the posterior 95\% credible bands under the proposed method.}
    \label{fig:comb_effect}
\end{figure}

Next, we examine whether we can recover the drug combination effect $\bm{h}(\cdot)$. 
We randomly selected one simulated dataset from 100 repeated simulations, and plotted the histogram of the true drug combination effects overlaid with the empirical density of the posterior expected combination effects in Figure S4 of the Supplementary Material.
As for individual-specific drug combination effects across visits, Figure \ref{fig:comb_effect} compares the simulation truths and the estimated combination effects for two randomly selected individuals in this dataset. Both Figure S4 and Figure \ref{fig:comb_effect} show that our model can well recover the drug combination effects.

For parameter estimation, Figure S6 in the Supplementary Material plots the 95\% estimated credible intervals (CI) for $\btbeta_{kq}$'s using the same simulated dataset, where the triangles represent the simulation truths. As shown in Figure S6, all the 95\% CI are centered around the simulated true values.  
As another metric of performance, we computed, for each simulated dataset, the mean squared error (MSE) taken as the averaged 
squared errors between the post-burn-in MCMC posterior samples and the simulated truth. Table S5 in the Supplementary Material summarizes the mean and standard deviation of MSE across 100 simulated datasets for $\btbeta_{kq}$'s. 
Both Table S5 and Figure S6 show that the proposed method performs well in terms of estimating the parameter values.
    
In addition, we compared the proposed method to two alternatives: the Normal+Linear and the DP+Linear methods. 
Figure \ref{fig:comb_effect} compares the estimated combination effects under the proposed model to those under the two alternative methods. The proposed method with the ddCRP prior and the ST kernel well recovered the ground truth, while both the Normal+Linear and DP+Linear methods had larger bias in estimating the drug combination effects. 

\begin{figure}[!htb]
    \centering
    \begin{tabular}{ccc}
    \includegraphics[width=.33\textwidth]{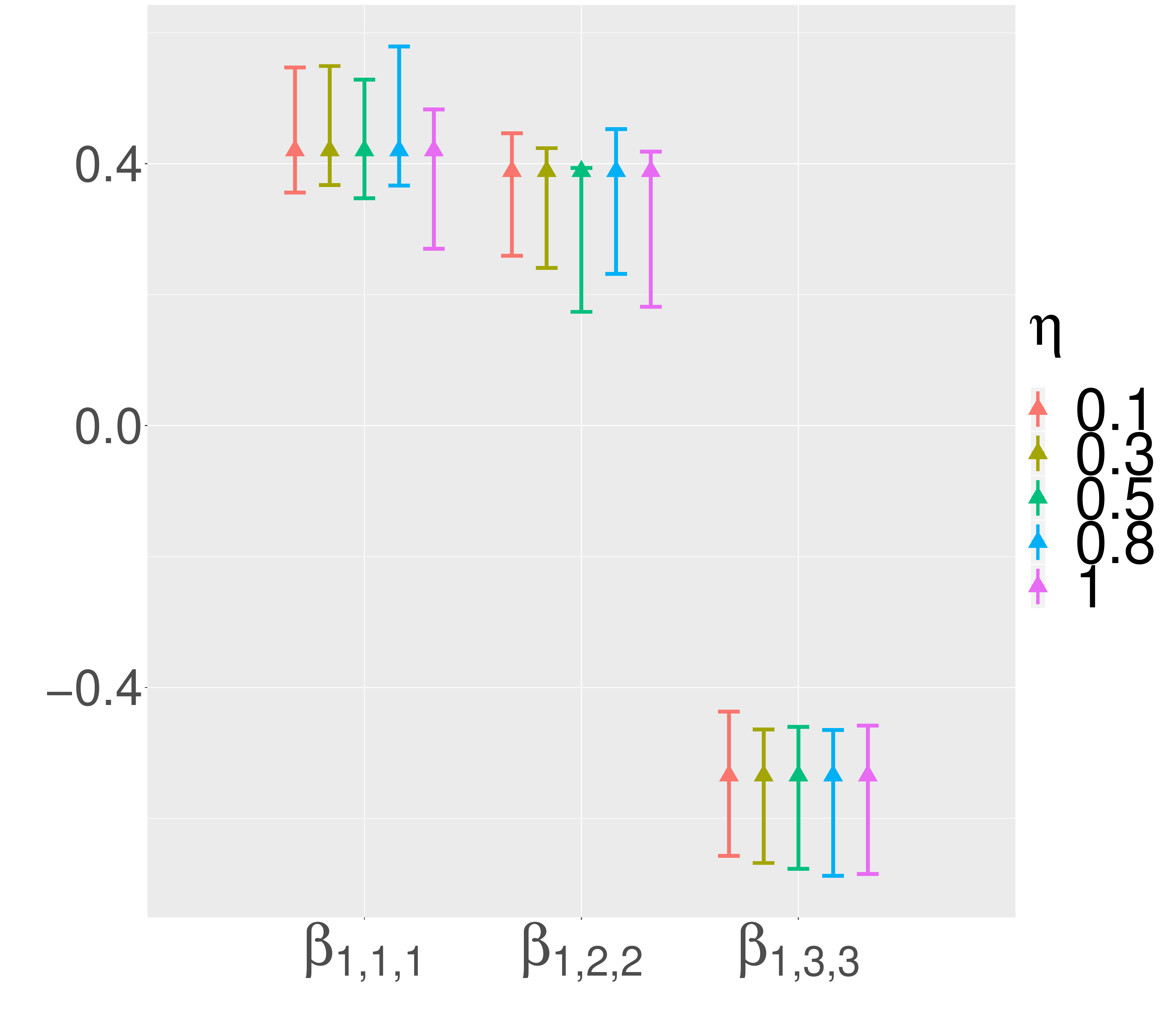} & 
    \includegraphics[width=.33\textwidth]{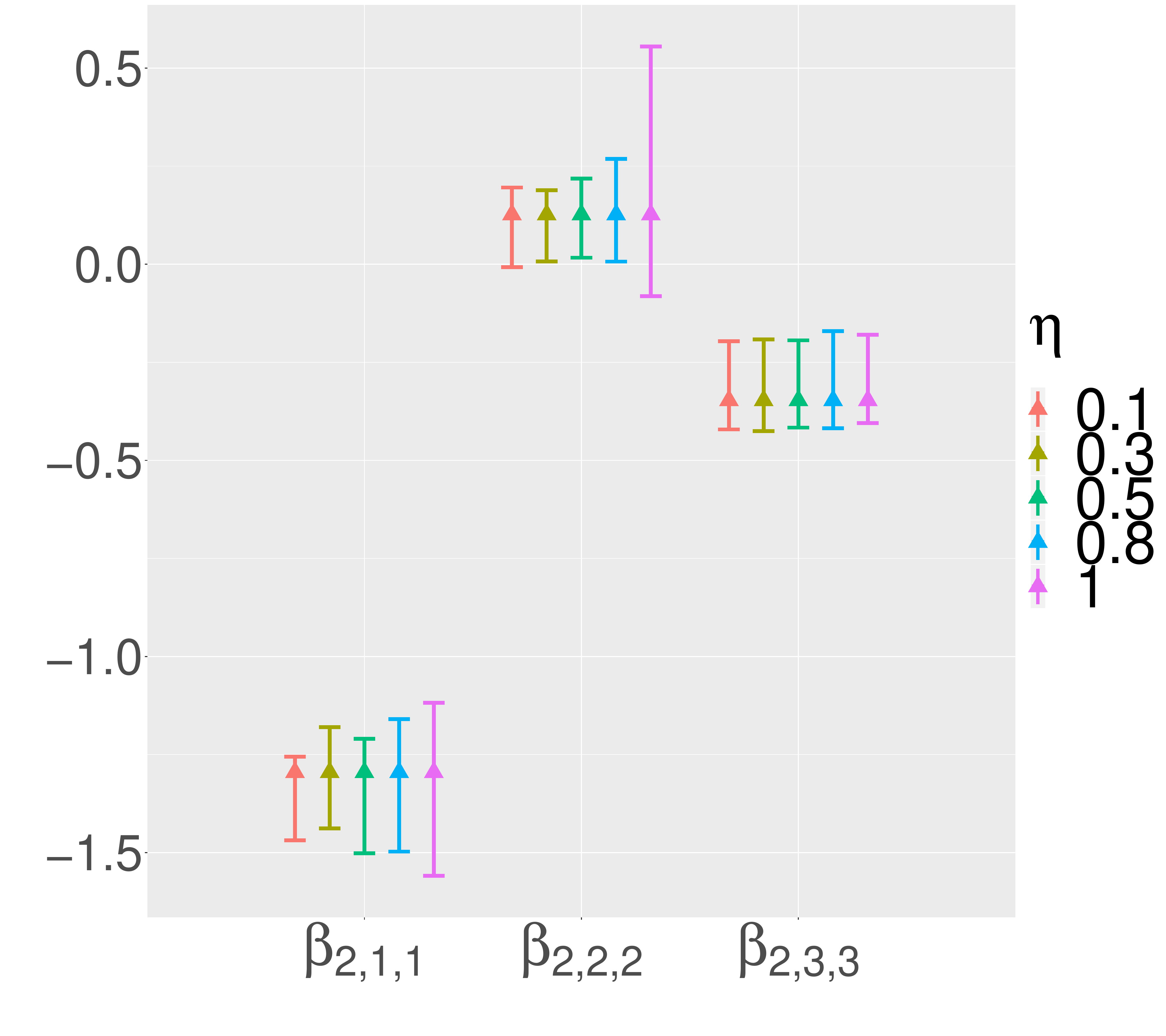} &
    \includegraphics[width=.33\textwidth]{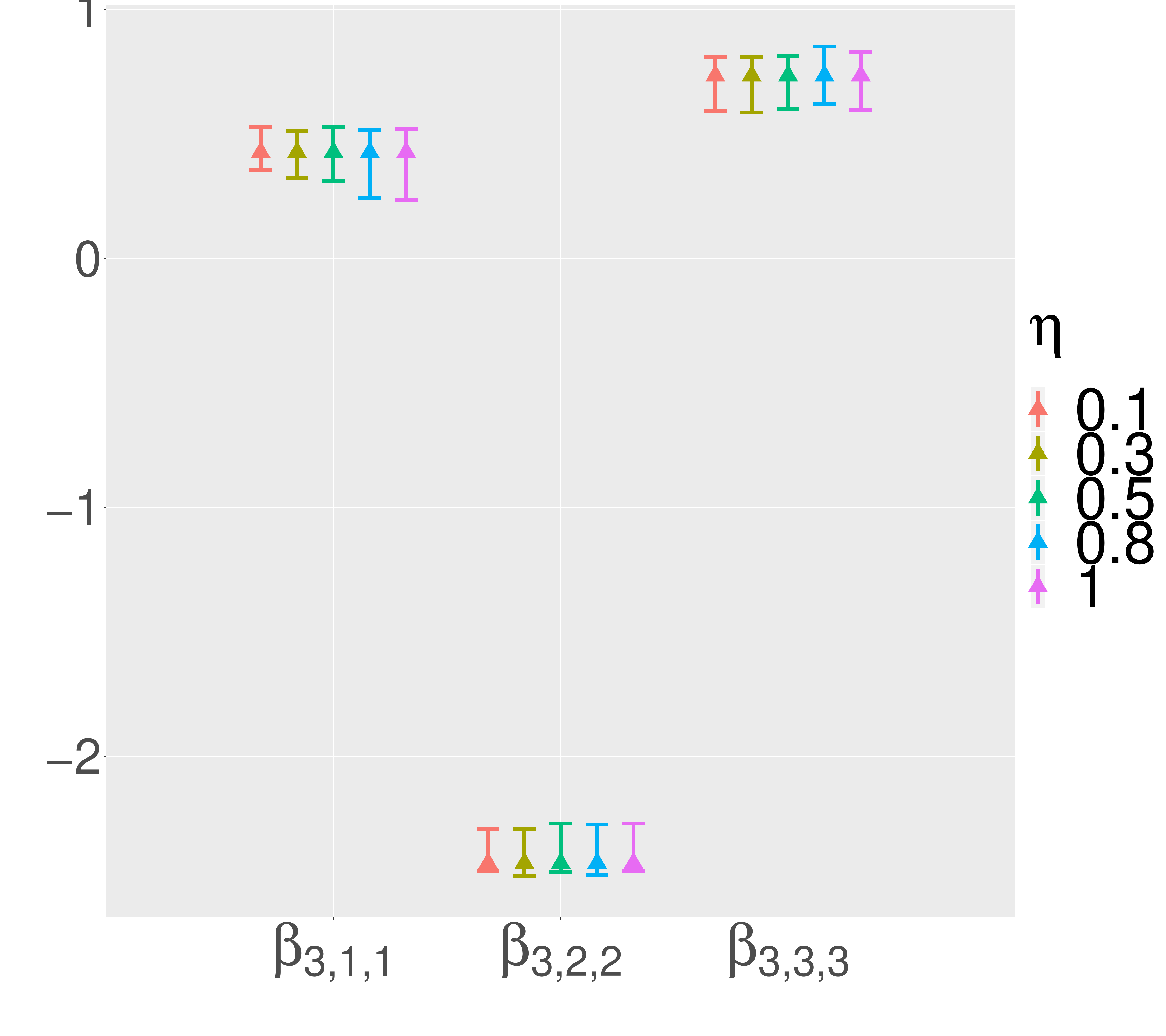} \\ 
    (a) Cluster $\#$1 & (b) Cluster $\#$2 & (c) Cluster $\#$3 \\
    \end{tabular}
    \caption{$95\%$ credible intervals for randomly selected $\{\tbeta_{kqs}\}_{k=1}^{r_n^o},_{q=1}^{Q},_{s=1}^{S}$ in the sensitivity analyses, where the triangles represent the simulated true values, and the colors represent different values of the decay factor $\eta$.}
    \label{fig:sensi_results}
\end{figure}

Lastly, to explore the sensitivity of the posterior inference with respect to the decay factor $\eta$, we conducted inference under several values of $\eta$ = 0.1, 0.3, 0.5, 0.8, 1 for one randomly selected simulated dataset. 
The decay factor $\eta$ was originally introduced in natural language processing to alleviate the peakiness of the ST kernel when the depth of the tree fragments is considerably large, in which case self similarities are disproportionately larger than similarities between two different trees. 
Therefore, the decay factor $\eta \in (0,1]$, which down-weights the contribution of large tree fragments to the kernel exponentially with their sizes, could have significant influence on the inference if the tree structure is deep. 
However, this is not the case in our application with relatively shallow trees. 
Figure \ref{fig:sensi_results} compares the parameter estimations under different values of $\eta$, showing that there is no significant difference among all these experiments. 

\section{Application: WIHS Data Analysis}
\label{sec:real}

The Women’s Interagency HIV Study (WIHS) is a multisite, longitudinal cohort study of women living with HIV and women at-risk for HIV in the United States \citep{barkan1998women, adimora2018cohort}.
Full details of the study design and prospective data collection are described at \url{https://statepi.jhsph.edu/wihs/wordpress}. Participants provide biological specimens, complete physical examinations, and undergo extensive assessment of demographic, clinical, and behavioral data via interviews at each visit.
Included in this assessment was the Center for Epidemiological Studies Depression Scale (CES-D, \citealt{radloff1977ces}), which is a self-report assessment of depressive symptoms spanning somatic (e.g., sleep and appetite difficulties), negative affect (e.g., loneliness and sadness), lack of positive affect (e.g., hopelessness), and interpersonal symptoms (e.g., people are unfriendly).
For the present analysis, we included all women from the Washington, D.C. site in the WIHS with at least five visits and complete CES-D data, which yielded $n=259$ individuals. We also extracted the following sociodemographic, behavioral, and clinical risk factors for depressive symptoms: age, race, smoking status, substance use (e.g., marijuana, cocaine, and heroin), body mass index (BMI), hypertension, CD4 count, and viral load. We selected $D=87$ representative ART regimens in \eqref{eqn:combined_drug_kernel} using the same criterion as in the simulation study. In particular, these representative ART regimens are combinations of 24 ART agents in five drug classes: NRTI, NNRTI, PI, INSTI, and EI. 

We applied the proposed model to the WIHS dataset using the same hyperparameters as in the simulation study and set the decay factor to be $\eta=0.5$. We performed the principal component analysis on the kernel weight matrix based on these 87 representative ART regimens, and selected the first $D^{\star}=45$ principal components that explain 99.9\% variation of the original matrix. 
We used 5,000 post burn-in samples after 5,000 iterations with a thinning factor of 10 for posterior inference. 
The proposed model identified three clusters, with the number of women in each cluster being 132, 84, and 43 respectively. 
Table S7 in the Supplementary Material summarizes the demographic, clinical, and behavioral characteristics of women in the three clusters at their initial visits, and Table S8 reports the frequency of the 24 ART agents and their corresponding drug classes used by women in the three clusters, respectively. 

\begin{figure}[!htb]
    \centering
    \begin{tabular}{cc}
    \includegraphics[width=.5\textwidth]{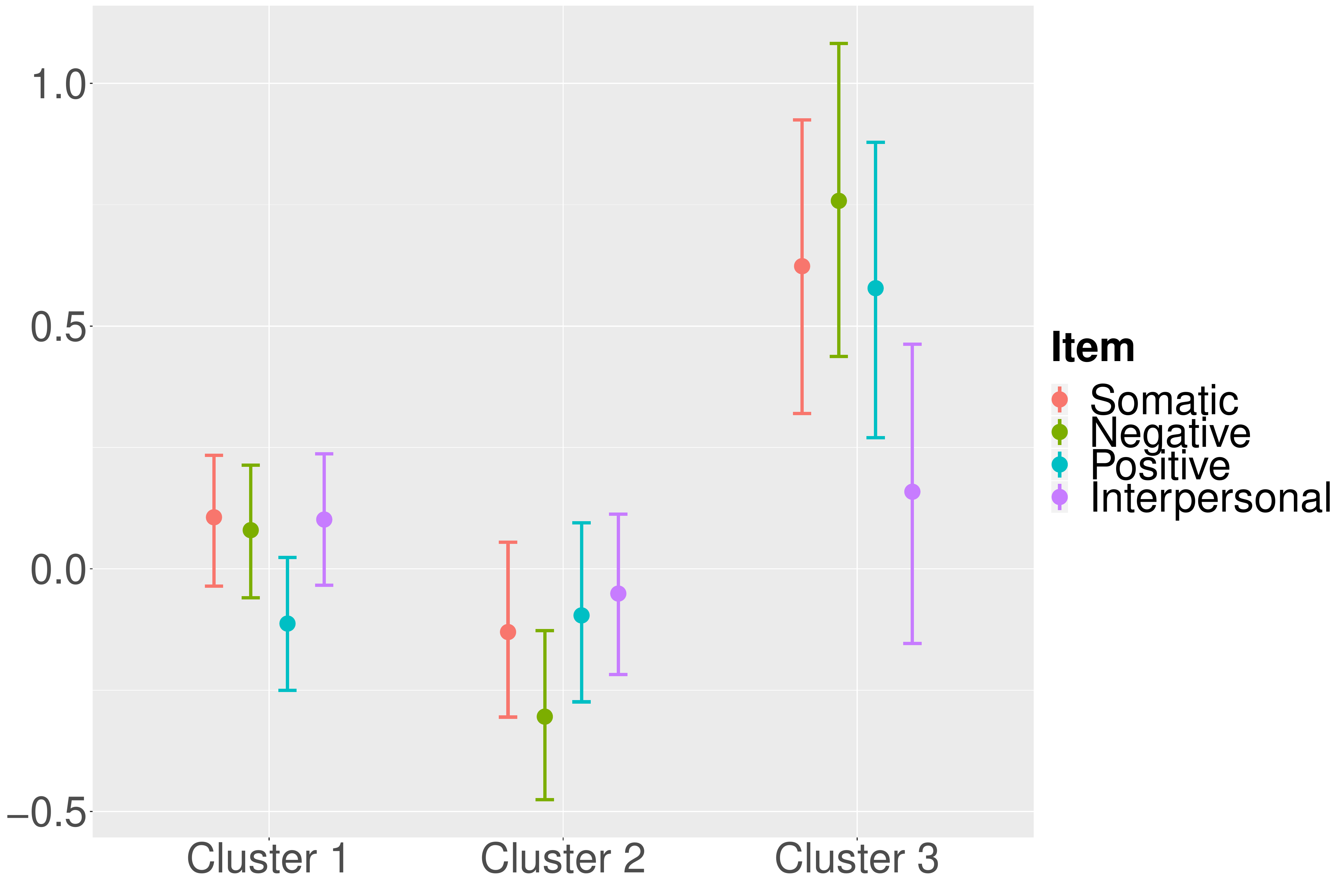} & 
    \includegraphics[width=.5\textwidth]{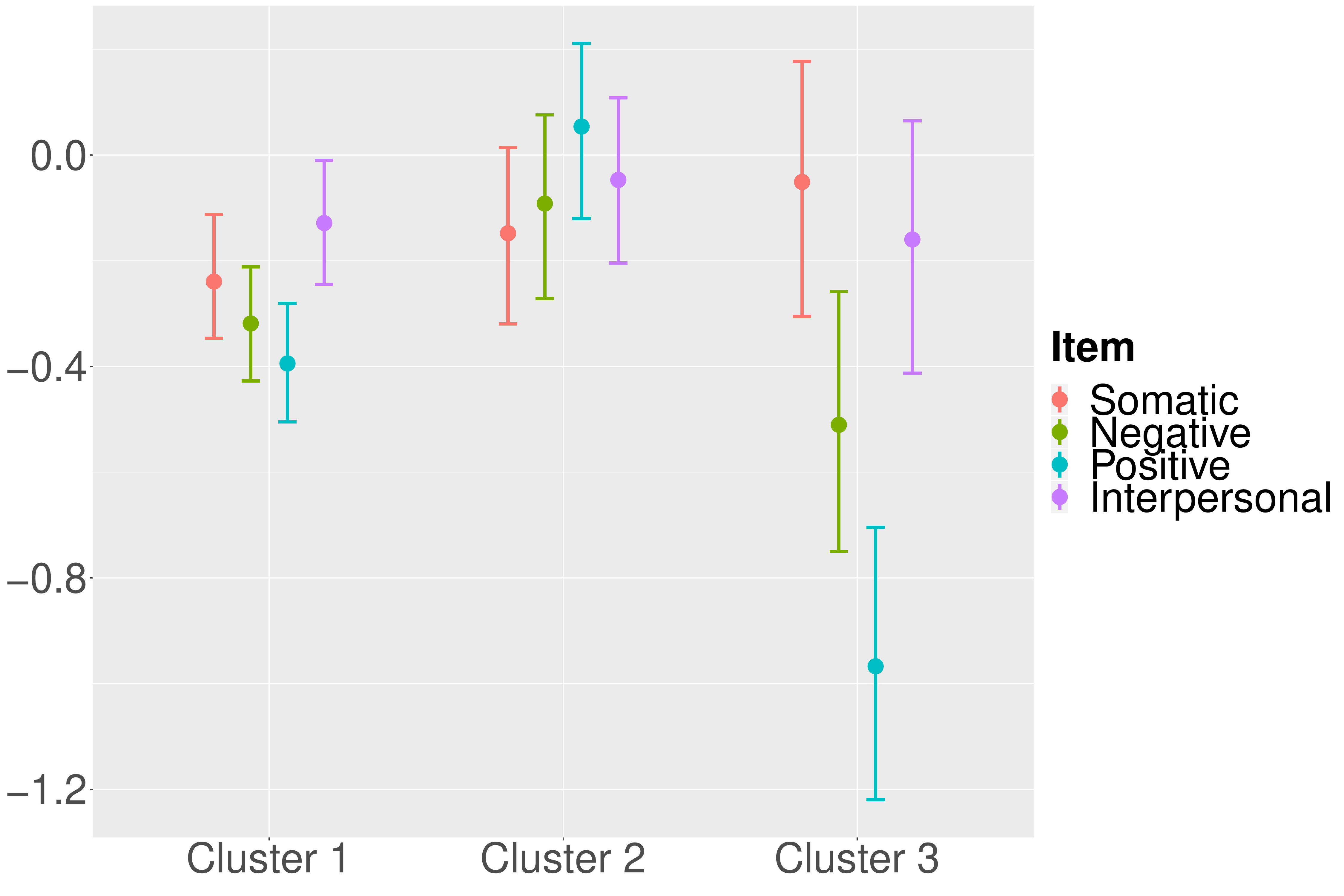} \\
    (a) Age & (b) CD4 count \\ 
    \includegraphics[width=.5\textwidth]{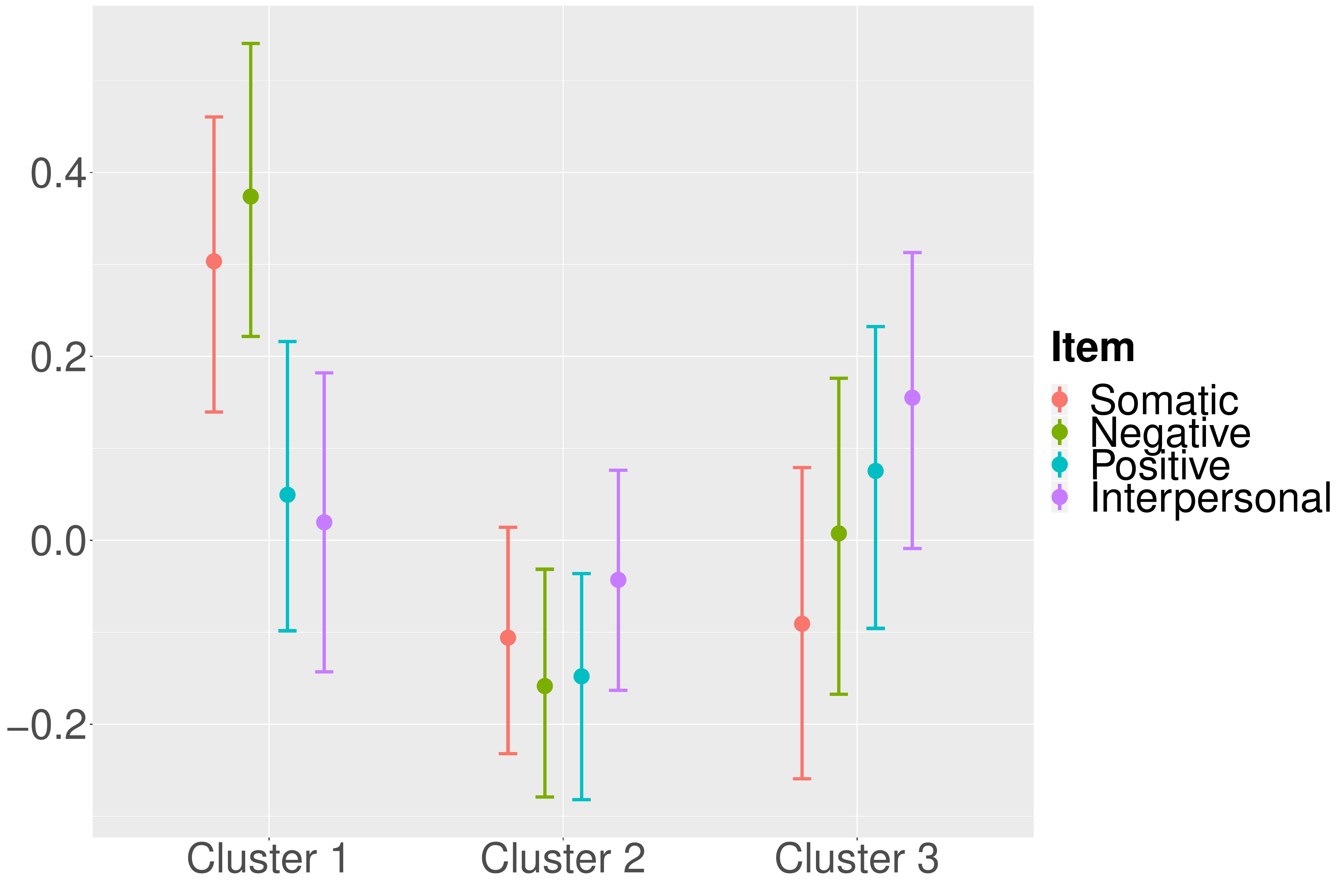} & 
    \includegraphics[width=.5\textwidth]{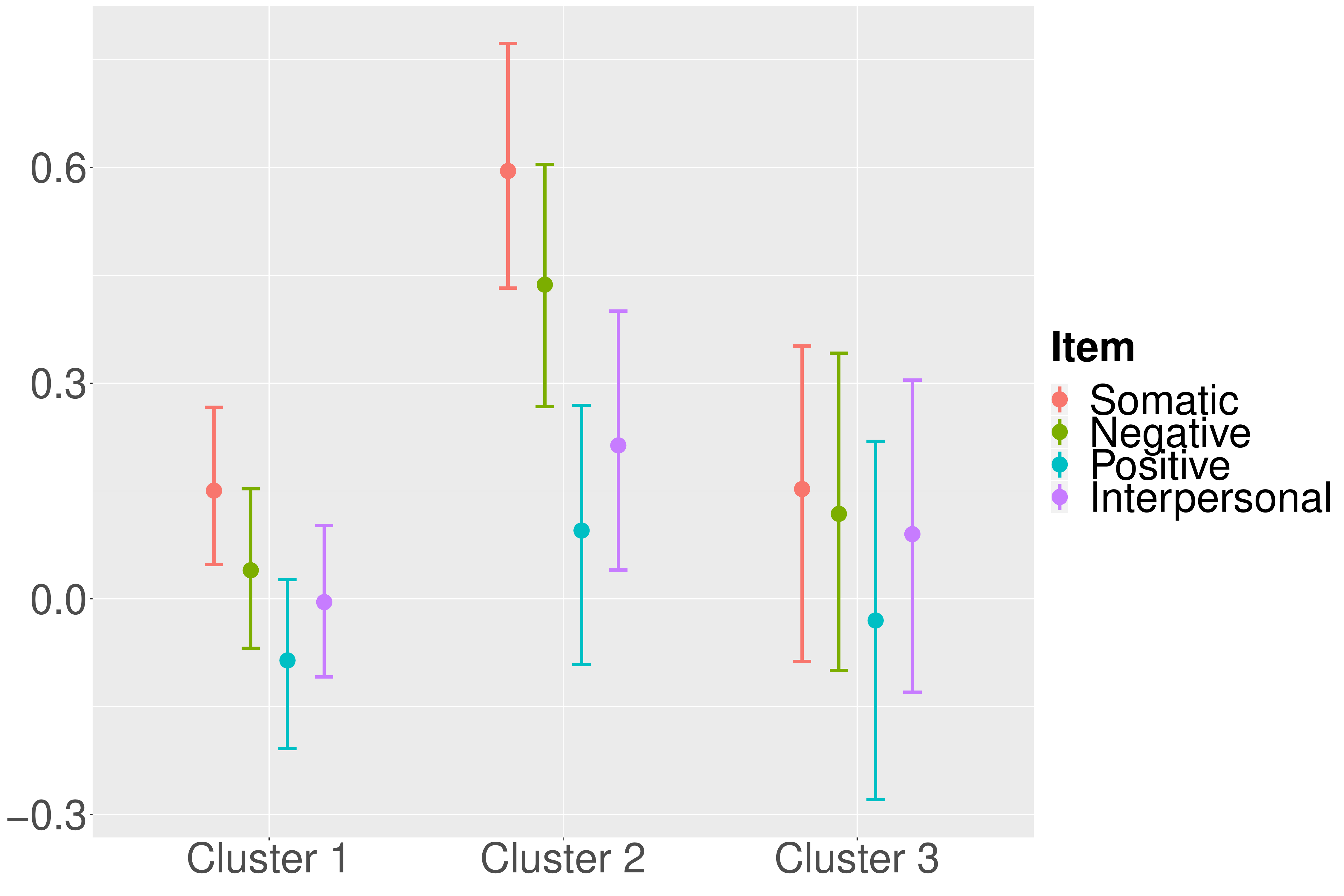} \\
    (c) Viral load & (d) Substance use \\
    \end{tabular}
   \caption{Posterior means and $95\%$ CIs for the estimated coefficients corresponding to age, CD4 count, viral load, and substance use in the real data analysis. The dots represent the posterior means and the colors indicate different depressive symptoms. }
    \label{fig:coef_esti}
\end{figure}

Figure \ref{fig:coef_esti} summarizes the posterior means and the corresponding $95\%$ credible intervals of the estimated coefficients with respect to age, CD4 count, viral load, and substance use on four depression items in each cluster.
As seen in Figure \ref{fig:coef_esti}, the effects of covariates on depressive symptoms were distinct among the three clusters. Panel (a) shows that younger people had higher depressive symptoms in cluster 2, but lower depressive symptoms in clusters 1 and 3.
Panels (b) and (c) indicate that higher CD4 and lower viral load are associated with lower depressive symptoms. Panel (d) shows a positive relationship between substance use and depressive symptoms. These findings are consistent with the literature \citep{berg2007behavioral, springer2009depression, grov2010loneliness, taniguchi2014depression}.

\begin{figure}[!htb]
    \centering
    \begin{tabular}{cc}
    \includegraphics[width=.5\textwidth]{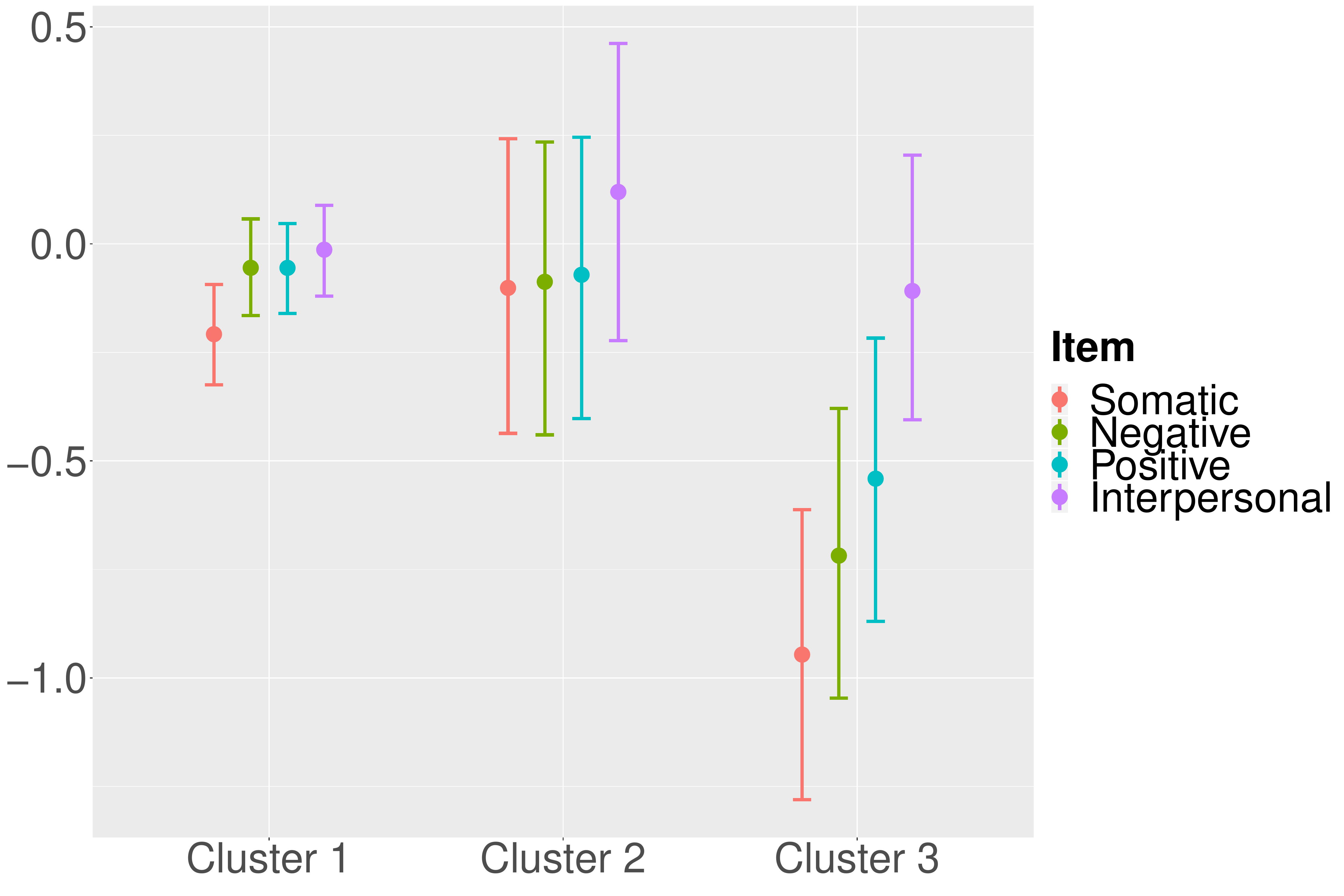} & 
    \includegraphics[width=.5\textwidth]{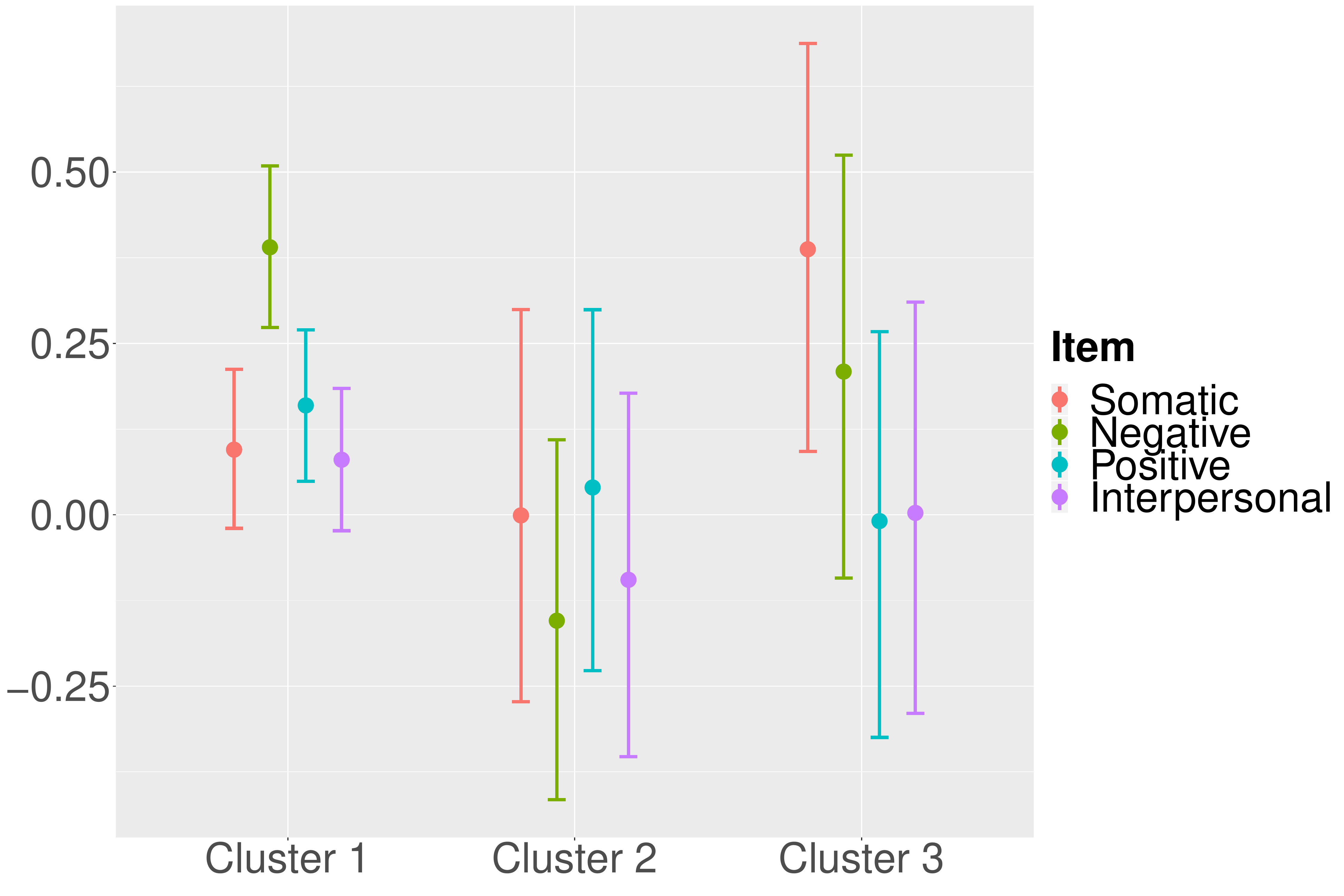} \\
    (a) Principal component $\#1$ & (b) Principal component $\#2$ \\ 
    \end{tabular}
   \caption{Posterior means and $95\%$ CIs for the estimated combination effects on four depressive symptoms with respect to the first two principal components in the WIHS  data analysis. The dots represent the posterior means and the colors indicate different depressive symptoms.}
    \label{fig:drug_effect_esti}
\end{figure}

Next, we report the effects of ART regimens, i.e., drug combinations, on depressive symptoms in each cluster. 
Figure \ref{fig:drug_effect_esti} plots the association between ART regimens and depressive symptoms with respect to the first two principal components in each cluster.
To explore the patterns and interpret the estimated drug combination effects, we further list the top five positively and negatively related ART regimens for each principal component in terms of the coefficients of the loading matrix in Table \ref{tab:drug_pc}.
As shown in Figure \ref{fig:drug_effect_esti}, the first principal component was negatively associated with all the depressive symptoms in cluster 1 and 3, but had little effects in cluster 2. 
In addition, the first principal component was positively associated with ART regimens consisting of two NRTI drugs FTC + TDF, an NNRTI drug EFV, RPV or NVP, and an additional INSTI drug RAL (Table \ref{tab:drug_pc}), which indicates a beneficial or protective effect for these ART regimens on depressive symptoms. 
In fact, combining two NRTI drugs as backbone with an additional NNRTI drug was recommended as one of the first-line therapies \citep{gunthard2014antiretroviral}, and previous clinical studies also reported that RAL was well-tolerated and provided desirable viral suppression when used with certain NRTIs such as TDF \citep{grinsztejn2007safety, markowitz2007rapid}. 
Conversely, negative relationships were observed between the first principle component and ART regimens consisting of two NRTI drugs AZT + LAM and a PI drug such as LPV, revealing worse depressive symptoms for women using these drug combinations. 
\cite{rabaud2005tolerability} reported that a large proportion of individuals receiving AZT + LAM + LPV experienced serious adverse effects, especially gastrointestinal side effects such as nausea and vomiting, leading to poor tolerability of this regimen and treatment discontinuation. 
Furthermore, the second principal component was positively associated with depressive symptoms in clusters 1 and 3. ART regimens consisting of two NRTI drugs AZT + LAM and an NNRTI drug such as EFV were positively related to the second principal component, while regimens consisting of two NRTI drugs FTC + TDF and two PI drugs such as ATZ + RTV were negatively related to the second principal component. 
Therefore, a combination of AZT, LAM, and EFV was estimated to have adverse effects on depressive symptoms whereas a combination of FTC, TDF, ATZ and RTV was estimated to have beneficial effects.
Indeed, \cite{gallant2006tenofovir} reported more frequent adverse effects and treatment discontinuation when individuals were on EFV combined with AZT and LAM instead of FTC and TDF. 
Conversely, adding PI drugs ATZ and RTV to NRTI drugs FTC and TDF yields both significant antiviral efficacy and safety \citep{soriano2011nevirapine}.

\begin{table}[!htb]
\centering
% \begin{adjustbox}{width=\columnwidth*9/10,center}
\begin{tabular}{ P{8cm} P{4cm} }
 \hline
 ART Regimens & Loading Coefficients \\
 \rowcolor{lightgray} \multicolumn{2}{c}{Principal Components $\#1$}  \\
 FTC + TDF + EFV & 0.182   \\
 FTC + TDF + RPV & 0.182   \\
 FTC + TDF + NVP & 0.181  \\
 FTC + TDF + EFV + RAL & 0.174 \\
 DDI + TDF + EFV & 0.162 \\
 ~\\
 AZT + LAM + LPV & -0.171  \\
 AZT + LAM + SQV & -0.165  \\
 DDI + LAM + LPV & -0.163  \\
 ABC + LAM + LPV & -0.162 \\
 AZT + LAM + IDV & -0.162  \\
 \rowcolor{lightgray} \multicolumn{2}{c}{Principal Components $\#2$ }  \\
 AZT + LAM + EFV & 0.179 \\
 AZT + LAM + NVP &  0.178 \\
 ABC + AZT + LAM + EFV & 0.169 \\
 ABC + AZT + EFV & 0.159 \\
 AZT + DDI + NVP & 0.156 \\
 ~\\
 FTC + TDF + DRV + RTV & -0.191 \\
 FTC + TDF + FPV + RTV & -0.190 \\
 FTC + TDF + ATZ + RTV & -0.189 \\
 DDI + TDF + ATZ + RTV & -0.182 \\
 FTC + TDF + ATZ & -0.175 \\
 \hline
\end{tabular}
% \end{adjustbox}
\caption{Top five positively and negatively related ART regimens for the first two principal components in terms of the coefficients of the loading matrix.}
\label{tab:drug_pc}
\end{table}

\begin{figure}[!htb]
    \centering
    \begin{tabular}{cc}
    \includegraphics[width=.5\textwidth]{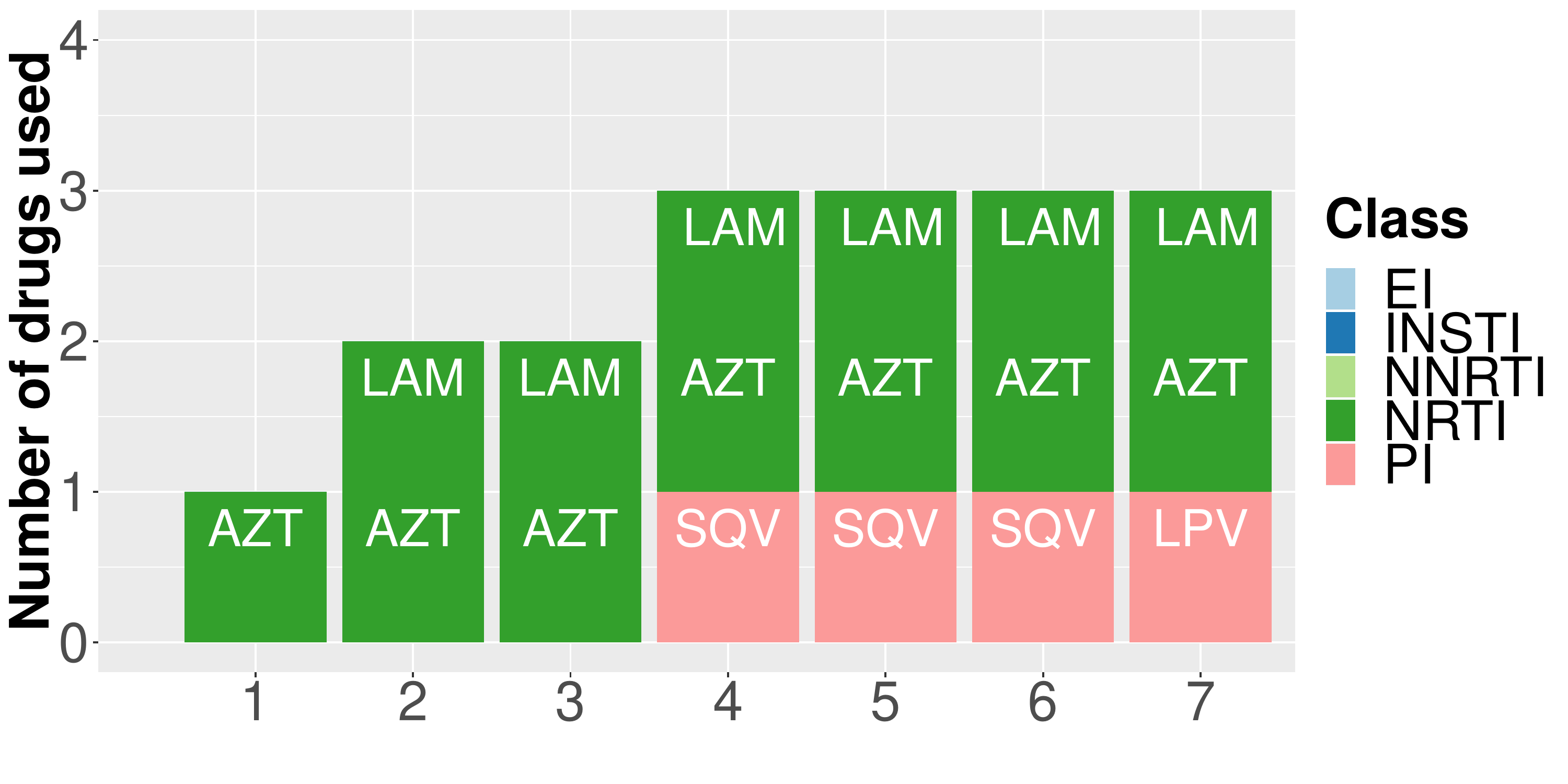} & 
    \includegraphics[width=.5\textwidth]{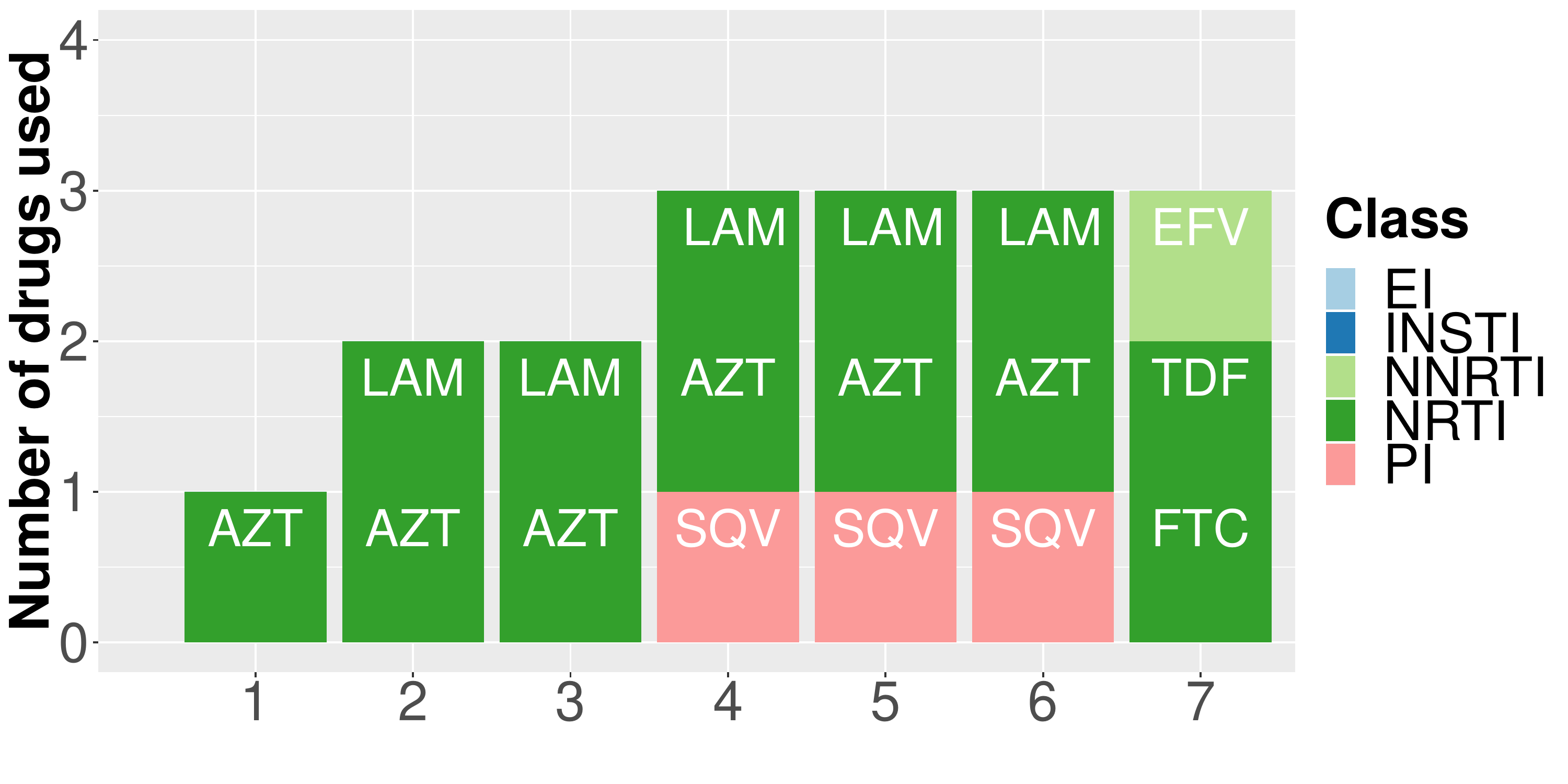} \\
	(a) ART use for scenario $\#$1 & (b) ART use for scenario $\#$2 \\ 
	\includegraphics[width=.5\textwidth]{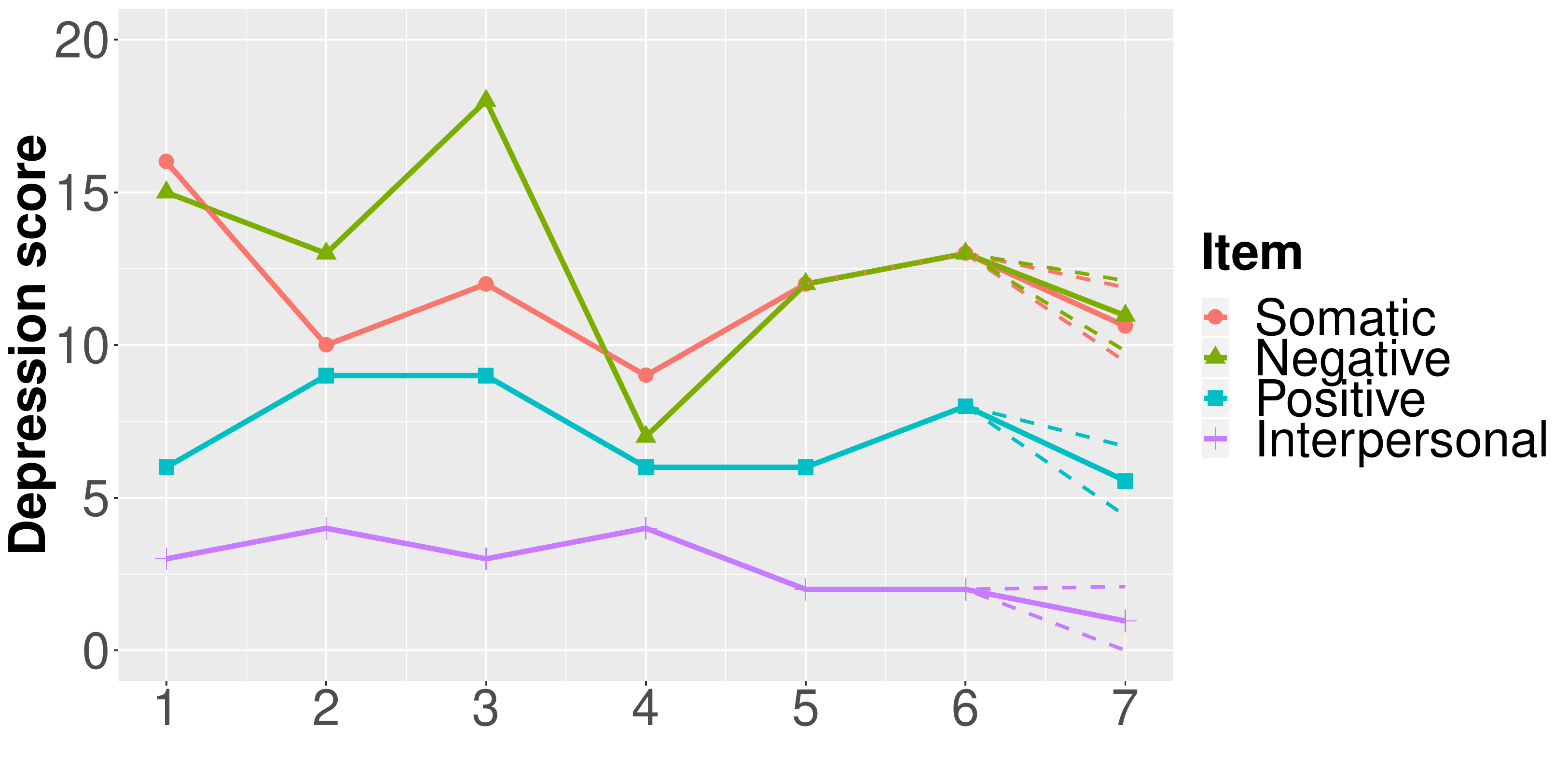} & 
	\includegraphics[width=.5\textwidth]{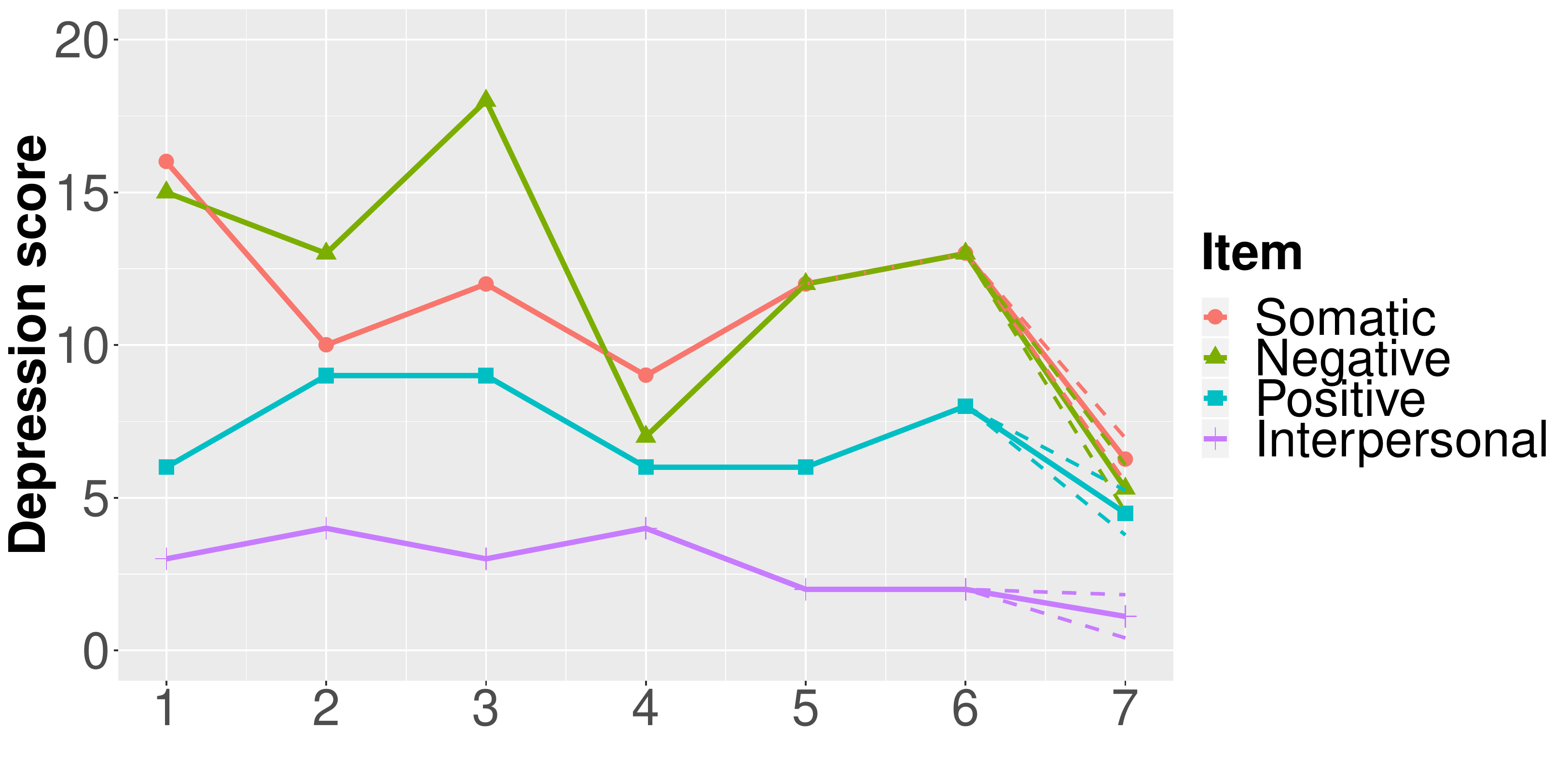} \\
    (c) Predictive depression score for scenario $\#$1 & (d) Predictive depression score for scenario $\#$2 \\
    \end{tabular}
    \caption{Predictive depression scores for an individual in the WIHS dataset with two different hypothetical scenarios of ART medication use. The dashed lines represent the predictive 95\% credible bands with respect to each depressive symptom. }
    \label{fig:dep_pred}
\end{figure}

The U.S. Department of Health and Human Services provides general guidelines on ART treatments; however, these guidelines do not take into account individual heterogeneity and treatment histories. 
To make clinical decisions tailored to each person (precision medicine), understanding the individualized adverse effect  of each possible drug combination will be one of the key contributors. 
The proposed method can accurately predict individuals' adverse effects of ART based on their clinical profiles, which can help guide clinicians to prescribe ART regimens.
For illustration, we randomly selected an individual from the WIHS dataset with seven visits in total, who started AZT (NRTI) at the first visit, added LAM (NRTI) at the second visit, and used the drug combination AZT + LAM + SQV (PI) from her fourth to sixth visits. 
Then we considered two hypothetical scenarios. 
In the first scenario, we assumed that the individual kept using the similar NRTI + PI drug combination as before but only replaced the PI drug SQV with a different PI drug LPV.
In the second scenario, this individual was switched to a distinct NRTI + NNRTI drug combination FTC (NRTI) + TDF (NRTI) + EFV (NNRTI).
Figure \ref{fig:dep_pred} plots the posterior predictive depression scores for this individual at the last visit based on the information from her previous six visits under the two hypothetical scenarios. 
As shown in Figure \ref{fig:dep_pred}(c), there were no significant differences between using different PI drugs when combined with NRTI drugs AZT and LAM as the backbone treatment. However, using NRTI drugs FTC and TDF as the backbone with NNRTI drug EFV demonstrated superior performance on alleviating depressive symptoms.
As a result, we would recommend the clinician to select the ART regimen FTC + TDF + EFV instead of AZT + LAM + LPV for this particular individual. This example demonstrates that the proposed method has the potential to guide more informed and effective personalized medicine in HIV clinical practice.   

To facilitate the implementation of the proposed method in the decision process of HIV clinicians, and for broad application in personalized medicine, we have created an interactive web application to illustrate this example using R package \pkg{shiny} \citep{changshiny2019}, available at \url{https://wjin.shinyapps.io/Rshiny/}. The web user interface interactively displays the predictive depression scores of an individual in response to the user's choice of the individual's clinical characteristics and ART medication use. Figure S9 in the Supplementary Material shows a screenshot of the web application.

\section{Conclusion}
\label{sec:dis}

To facilitate a precision medicine approach, we proposed a novel Bayesian nonparametric approach to estimate the effects of ART regimens on depressive symptoms. The method is built upon the ST kernel method that quantifies similarities among ART regimens and the ddCRP that accounts for individuals' heterogeneity in both treatment histories and clinical characteristics. Through simulation studies and analysis of the WIHS dataset, we have demonstrated that the proposed model can accurately estimate the drug combination effects and yield meaningful and interpretable results. 

There are several potential extensions. First, the current similarity score is parameterized by a hyperparameter $\eta$. We could impose a prior on $\eta$ and estimate it from the posterior inference. It will require us to develop more efficient posterior samplers because in each iteration of MCMC the similarity matrix needs to be recalculated at the current value of $\eta$. 
Second, the similarity between ART regimens may also depend on the individuals' socio-demographic, behavioral, and clinical characteristics. We could extend the model to account for these factors by modifying the parameter ${\bm \gamma}_{id}$ in \eqref{eqn:combined_drug_kernel} as a function of these variables. 
Finally, combination therapies are needed for many complex diseases beyond HIV such as cancer and chronic diseases. Each chronic condition requires long-term medication use. The proposed method can be applied to such electronic health records datasets \citep{gill2010antidepressant} to examine the side effects of combination therapies, potentially yielding better therapy management for the elderly population and which has the potential to reduce public healthcare costs. 

\section*{Acknowledgment}
This work was supported by the Johns Hopkins University Center for AIDS Research NIH/NIAID fund (P30AI094189) 2019 faculty development award to Dr. Xu, NSF 1940107 to Dr. Xu,  NSF DMS1918854 to Drs. Xu and Rubin, and NSF DMS1918851 to Dr. Ni.

\bibliographystyle{apalike}
\bibliography{Reference}

\newpage
\begin{center}
\textbf{\large Supplemental Materials for 
``A Bayesian Nonparametric Approach for Inferring Drug Combination Effects on Mental Health in People with HIV"}
\end{center}

\section*{A: Details of MCMC}

\subsection*{A1: Summary of Model}

\begin{align*}
   \bm{Y}_{ij} \mid \bm{\theta}_i 
            & \sim \mathcal{N}\left(\bm{\beta}_{i}\bm{X}_{ij} + \bm{h}(Z_{ij}) + \bm{\omega}_{ij}, 
                \sigma_{\varepsilon}^2\bm{I}_{Q}\right),  \quad i=1,2,\dots,n, \quad j=1,2,\dots, J_i, \\
   \bm{\theta}_i & = (\bbeta_{i}, \bgamma^{\star}_{i}) \sim \text{ddCRP}(m_0, s, G_0),
\end{align*}
where $\bm{h}(Z_{ij}) = \frac{\sum_{d=1}^{D} \kappa(Z_{ij}, z_d)\bm{\gamma}_{id}}{\sum_{d=1}^{D} \kappa(Z_{ij}, z_d)}$ is the drug combination effect that will be estimated by a principal component regression, the covariates of which are derived from the first $D^{\star}$ principal components of the kernel weight matrix.
Let $\bgamma^{\star}_{i}$ be the corresponded coefficients for the principal component regression, and $\btth_k=(\btbeta_{k}, \btgamma^{\star}_{k})$'s be the $r_n$ unique values of $\bm{\theta}_i$ induced by ddCRP, where $k=1,2,\dots,r_n$.

We complete the model by assuming $m_0\sim \text{Gamma}(c_0,d_0)$, $\sigma_{\varepsilon}^2 \sim \text{Inverse-Gamma}(g_1, g_2)$, $\btbeta_{kq} \sim \mathcal{N}(\bm{e}_q,\bm{B}_q)$ and $\btgamma^{\star}_{kq} \sim \mathcal{N}(\bm{f}_q,\bm{\Lambda}_q)$ independently for $k=1,2,\dots,r_n$, and $q=1,2,\dots,Q$, where $\bm{e}_q \sim \mathcal{N}(\bm{0},\bm{E}_0)$, $\bm{B}_q \sim \text{Inverse-Wishart}(b_0, \bm{B}_0^{-1})$, $\bm{f}_q \sim \mathcal{N}(\bm{0},\bm{F}_0)$, $\bm{\Lambda}_q \sim \text{Inverse-Wishart}(\lambda_0, \bm{\Lambda}_0^{-1})$, $\bm{\omega}_{ij} \sim \mathcal{N}(\bm{0}, \sigma_{\varepsilon}^2 \bm{\Sigma}_{\omega})$, and $p(\bm{\Sigma_\omega}) \propto \det(\bm{\Sigma_\omega})$.

\subsection*{A2: Posterior Computation}    

\subsubsection*{A2.1: Update $\pi_n$, the partition induced by ddCRP}
\begin{align*}
    p(i \in S_k^{-}\mid \cdot) \propto p(\pi_n^{i\to k} \mid m_0, s, \bm{\sigma})\prod_{j=1}^{J_i}p(\bm{Y}_{ij} \mid \btth_k)
\end{align*}
for $i=1,2,\dots,n$ and $k=0,1,\dots,r_n$, where $\btth_0$ is a new and independent draw from $G_0$.

\subsubsection*{A2.2: Update $m_0$, the mass parameter of ddCRP}
\begin{align*}
    p(m_0\mid \cdot) \propto p(m_0)p(\pi_n \mid m_0, s, \bm{\sigma})
    \propto p(m_0)\frac{\Gamma(m_0)}{\Gamma(m_0+n)}m_0^{r_n}.
\end{align*}
Following the idea of \cite{escobar1995bayesian}, we assign a Gamma distribution prior on $m_0$ and introduce an auxiliary variable $\tau_0 \sim \text{Beta}(m_0+1, n)$. Then a closed form full conditional posterior will be available for the mass parameter $m_0$, which is a mixture of Gamma distributions, i.e., 
\begin{align*}
    \tau_0 \mid m_0, \cdot & \sim \text{Beta}(m_0+1, n) \\
    m_0 \mid \tau_0, \cdot & \sim \frac{c_0+r_n-1}{c_0+r_n-1+n(d_0-\log(\tau_0))} \text{Gamma}(c_0+r_n,d_0-\log(\tau_0)) \\
    & \quad + \frac{n(d_0-\log(\tau_0))}{c_0+r_n-1+n(d_0-\log(\tau_0))} \text{Gamma}(c_0+r_n-1,d_0-\log(\tau_0)).
\end{align*}

\subsubsection*{A2.3: Update $\bm{\sigma}$, the permutation of subjects}
We use Metropolis–Hastings algorithm to update $\bm{\sigma}$. We first propose a new permutation $\bm{\sigma}^*$ by shuffling some randomly chosen integers in the current permutation $\bm{\sigma}$ and leaving the rest in their current positions. As a symmetric proposal distribution, the proposed new $\bm{\sigma}^*$ is accepted with probability
\begin{align*}
    \min\left\{\frac{p(\pi_n\mid m_0, s, \bm{\sigma}^*)p(\bm{\sigma}^*)}
    {p(\pi_n\mid m_0, s, \bm{\sigma})p(\bm{\sigma})}, 1\right\}
    =  \min\left\{\frac{p(\pi_n\mid m_0, s, \bm{\sigma}^*)}{p(\pi_n\mid m_0, s, \bm{\sigma})}, 1\right\},
\end{align*}
since we are assuming a uniform distribution prior on the permutation.

\subsubsection*{A2.4: Update $\{\btth_k=(\btbeta_{k}, \btgamma^{\star}_{k})\}_{k=1}^{r_n}$, the cluster specific parameters}
\begin{align*}
    p(\btth_k \mid \bm{Y}_i: i\in S_k) 
    \propto p(\btth_k)\prod_{i\in S_k}\prod_{j=1}^{J_i}p(\bm{Y}_{ij} \mid \btth_k) 
    \propto G_0 \prod_{i\in S_k}\prod_{j=1}^{J_i}\mathcal{N}(\bm{Y}_{ij}; \btbeta_{k}\bm{X}_{ij} 
    + \btgamma^{\star}_{k}\tilde{\bm{H}}_{ij} + \bm{\omega}_{ij}, \sigma_{\varepsilon}^2\bm{I}_{Q}).
\end{align*}

\begin{itemize}
    \item Update $\btbeta_{kq}$, $k=1,2,\dots,r_n$, $q=1,2,\dots, Q$
        \begin{align*}
           p(\btbeta_{kq} \mid \cdot) 
           & \propto  \mathcal{N}(\bm{e}_{q},\bm{B}_{q})\prod_{i\in S_k}\prod_{j=1}^{J_i}
           \mathcal{N}(Y_{ijq}; \bm{X}_{ij}^{T}\btbeta_{kq} + \tilde{\bm{H}}_{ij}^{T}\btgamma^{\star}_{kq} 
           + \omega_{ijq}, \sigma_{\varepsilon}^2) \propto \mathcal{N}(\bm{\mu}_n,\bm{V}_n),
        \end{align*}
    where $\bm{\mu}_{n} = \bm{V}_{n}\left(\frac{1}{\sigma_{\varepsilon}^2}
    \sum_{i\in S_k}\sum_{j=1}^{J_i}\tilde{Y}_{ijq}\bm{X}_{ij}+\bm{B}_{q}^{-1}\bm{e}_{q}\right)$, 
    $\bm{V}_{n}^{-1} = \frac{1}{\sigma_{\varepsilon}^2}\sum_{i\in S_k}\sum_{j=1}^{J_i}\bm{X}_{ij}\bm{X}_{ij}^{T}+\bm{B}_{q}^{-1}$, and $\tilde{Y}_{ijq}=Y_{ijq}-\tilde{\bm{H}}_{ij}^{T}\btgamma^{\star}_{kq}-\omega_{ijq}$ if $i\in S_k$.

    \item Update $\btgamma^{\star}_{kq}$, $k=1,2,\dots,r_n$, $q=1,2,\dots, Q$ 
      \begin{align*}
        p(\btgamma^{\star}_{kq} \mid \cdot) 
        & \propto  \mathcal{N}(\bm{f}_{q},\bm{\Lambda}_{q})\prod_{i\in S_k}\prod_{j=1}^{J_i}
        \mathcal{N}(Y_{ijq}; \bm{X}_{ij}^{T}\btbeta_{kq} + \tilde{\bm{H}}_{ij}^{T}\btgamma^{\star}_{kq} 
        + \omega_{ijq}, \sigma_{\varepsilon}^2) \propto \mathcal{N}(\bm{\mu}_n,\bm{V}_n),
        \end{align*}
    where $\bm{\mu}_{n} = \bm{V}_{n}\left(\frac{1}{\sigma_{\varepsilon}^2}
    \sum_{i\in S_k}\sum_{j=1}^{J_i}\tilde{Y}_{ijq}\tilde{\bm{H}}_{ij}+\bm{\Lambda}_{q}^{-1}\bm{f}_{q}\right)$, 
    $\bm{V}_{n}^{-1} = \frac{1}{\sigma_{\varepsilon}^2}\sum_{i\in S_k}\sum_{j=1}^{J_i}\tilde{\bm{H}}_{ij}\tilde{\bm{H}}_{ij}^{T}
    +\bm{\Lambda}_{q}^{-1}$, and $\tilde{Y}_{ijq}=Y_{ijq}-\bm{X}_{ij}^{T}\btbeta_{kq}-\omega_{ijq}$ if $i\in S_k$.
\end{itemize}

\subsubsection*{A2.5: Update $\{\bm{e}_{q}\}_{q=1}^{Q}$, $\{\bm{B}_{q}\}_{q=1}^{Q}$, $\{\bm{f}_{q}\}_{q=1}^{Q}$, $\{\bm{\Lambda}_{q}\}_{q=1}^{Q}$, the hyper-parameters}
\begin{itemize}
    \item Update $\bm{e}_{q}$, $q=1,2,\dots, Q$
        \begin{align*}
            p(\bm{e}_{q}\mid \cdot) 
            \propto \mathcal{N}(\bm{e}_0,\bm{E}_0) 
            \prod_{k=1}^{r_n}\mathcal{N}(\btbeta_{kq}; \bm{e}_{q}, \bm{B}_{q})
            \propto \mathcal{N}(\bm{\mu}_n,\bm{V}_n),
        \end{align*}
        where $\bm{\mu}_n = \bm{V}_n\left(\bm{E}_0^{-1}\bm{e}_0+\bm{B}_{q}^{-1}\sum_{k=1}^{r_n}\btbeta_{kq}\right)$ and
        $\bm{V}_n^{-1} = \bm{E}_{0}^{-1}+r_n \bm{B}_{q}^{-1}$.

    \item Update $\bm{B}_{q}$, $q=1,2,\dots, Q$   
        \begin{align*}
            p(\bm{B}_{q}\mid \cdot)
            & \propto \text{Inverse-Wishart}(b_0, \bm{B}_0^{-1})\prod_{k=1}^{r_n}
            \mathcal{N}(\btbeta_{kq}; \bm{e}_{q}, \bm{B}_{q}) 
            \propto \text{Inverse-Wishart}(b_n, \bm{B}_n^{-1}),
        \end{align*} 
        where $b_n=b_0+r_n$ and $\bm{B}_n=\bm{B}_0+\sum_{k=1}^{r_n}(\btbeta_{kq}-\bm{e}_{q})(\btbeta_{kq}-\bm{e}_{q})^T$.

    \item Update $\bm{f}_{q}$, $q=1,2,\dots, Q$
        \begin{align*}
            p(\bm{f}_{q}\mid \cdot) 
            \propto \mathcal{N}(\bm{f}_0,\bm{F}_0) 
            \prod_{k=1}^{r_n}\mathcal{N}(\btgamma^{\star}_{kq}; \bm{f}_{q}, \bm{\Lambda}_{q})
            \propto \mathcal{N}(\bm{\mu}_n,\bm{V}_n),
        \end{align*}
        where $\bm{\mu}_n = \bm{V}_n\left(\bm{F}_0^{-1}\bm{f}_0+\bm{\Lambda}_{q}^{-1}\sum_{k=1}^{r_n}\btgamma_{kq}^{\star}\right)$ and
        $\bm{V}_n^{-1} = \bm{F}_{0}^{-1}+r_n \bm{\Lambda}_{q}^{-1}$.

    \item Update $\bm{\Lambda}_{q}$, $q=1,2,\dots, Q$   
        \begin{align*}
            p(\bm{\Lambda}_{q}\mid \cdot)
            & \propto \text{Inverse-Wishart}(\lambda_0, \bm{\Lambda}_0^{-1})\prod_{k=1}^{r_n}
            \mathcal{N}(\btgamma^{\star}_{kq}; \bm{f}_{q}, \bm{\Lambda}_{q}) 
            \propto \text{Inverse-Wishart}(\lambda_n, \bm{\Lambda}_n^{-1}),
        \end{align*} 
        where $\lambda_n=\lambda_0+r_n$ and $\bm{\Lambda}_n=\bm{\Lambda}_0+\sum_{k=1}^{r_n}(\btgamma^{\star}_{kq}-\bm{f}_{q})(\btgamma^{\star}_{kq}-\bm{f}_{q})^T$.
\end{itemize}

\subsubsection*{A2.6: Update $\{\bm{\omega}_{ij}\}_{i=1}^{n},_{j=1}^{J_i}$, the normal correlation term} 
  \begin{align*}
    p(\bm{\omega}_{ij}\mid \cdot) 
    & \propto \mathcal{N}(\bm{0}, \sigma_{\varepsilon}^2\bm{\Sigma}_{\omega})
    \mathcal{N}(\bm{Y}_{ij}; \bbeta_{i}\bm{X}_{ij} + \bgamma^{\star}_{i}\tilde{\bm{H}}_{ij} + \bm{\omega}_{ij}, 
    \sigma_{\varepsilon}^2\bm{I}_{Q}) \propto \mathcal{N}(\bm{\mu}_n,\bm{V}_n),
  \end{align*}
  where $\bm{\mu}_{n} = \bm{V}_{n}\left(\frac{1}{\sigma_{\varepsilon}^2}\tilde{\bm{Y}}_{ij}\right)$, 
  $\bm{V}_{n}^{-1} = \frac{1}{\sigma_{\varepsilon}^2}\bm{I}_{Q} + \frac{1}{\sigma_{\varepsilon}^2}\bm{\Sigma}_{\omega}^{-1}$, and
  $\tilde{\bm{Y}}_{ij}=\bm{Y}_{ij}-\bbeta_{i}\bm{X}_{ij}-\bgamma^{\star}_{i}\tilde{\bm{H}}_{ij}$.

\subsubsection*{A2.7: Update $\bm{\Sigma}_{\omega}$, the correlation matrix} 
  \begin{align*}
    p(\bm{\Sigma}_{\omega} \mid \cdot) \propto \text{det}(\bm{\Sigma}_{\omega})
      \prod_{i=1}^{n}\prod_{j=1}^{J_i}\mathcal{N}\left(\bm{\omega}_{ij}; \bm{0}, \sigma_{\varepsilon}^2\bm{\Sigma}_{\omega}\right).
  \end{align*}   
  Since there is no closed-form solution, we will update it by Metropolis-Hasting algorithm.  

\subsubsection*{A2.8: Update $\sigma_{\varepsilon}^2$, the variance of i.i.d normal errors}     
{\small 
 \begin{align*}
    p(\sigma_{\varepsilon}^2\mid \cdot) 
    \propto \text{Inverse-Gamma}(g_1, g_2)
    \prod_{i=1}^{n}\prod_{j=1}^{J_i}\prod_{q=1}^{Q}
    \mathcal{N}(Y_{ijq}; \bm{X}_{ij}^{T}\btbeta_{iq} + \tilde{\bm{H}}_{ij}^{T}\btgamma^{\star}_{iq} 
    + \omega_{ijq}, \sigma_{\varepsilon}^2) 
    \propto \text{Inverse-Gamma}(g_1^*, g_2^*),
  \end{align*}
where $g_1^* = g_1 + \frac{1}{2}\sum\limits_{i=1}^{n}\sum\limits_{j=1}^{J_i}\sum\limits_{q=1}^{Q}1$, and $g_2^* = g_2 + \frac{1}{2}\sum\limits_{i=1}^{n}\sum\limits_{j=1}^{J_i}\sum\limits_{q=1}^{Q}(Y_{ijq}-\bm{X}_{ij}^{T}\btbeta_{iq}-\tilde{\bm{H}}_{ij}^{T}\btgamma^{\star}_{iq}-\omega_{ijq})^2$.}

\section*{B: Supplementary Tables and Figures}

% Table S1: simulation truths
\begin{table}[!htb]
\centering
\begin{tabular}{c|c}
  \hline
  Parameters & Simulation Truths  \\ 
  \hline
  $\btbeta_{11}$ & (0.4201738, -1.5065858,  0.4573016) \\
  $\btbeta_{12}$ & (0.1002570,  0.3885576, -2.5187332) \\
  $\btbeta_{13}$ & (0.8705657, -0.3111586, -0.5348084) \\
  
  $\btbeta_{21}$ & (-1.2951632, -0.07094494, -0.7004121) \\
  $\btbeta_{22}$ & (-0.8044954,  0.12646919, -0.3280640) \\
  $\btbeta_{23}$ & (1.3418530, -0.98949773, -0.3472228) \\
  
  $\btbeta_{31}$ & (0.4265138, -0.2214469,  0.1368007) \\
  $\btbeta_{32}$ & (-0.3282160, -2.4289411, -0.5135745) \\
  $\btbeta_{33}$ & (0.5458084,  1.7959664,  0.7342632) \\
  \hline
\end{tabular}
\caption*{Table S1: Simultion truths of the parameters $\{\btbeta_{kq}\}_{k=1}^{r_n^o},_{q=1}^{Q}$.}
\end{table}

\newpage
% Figure S2: acf plot
\begin{figure}[!htb]
    \centering
    \includegraphics[width=\columnwidth]{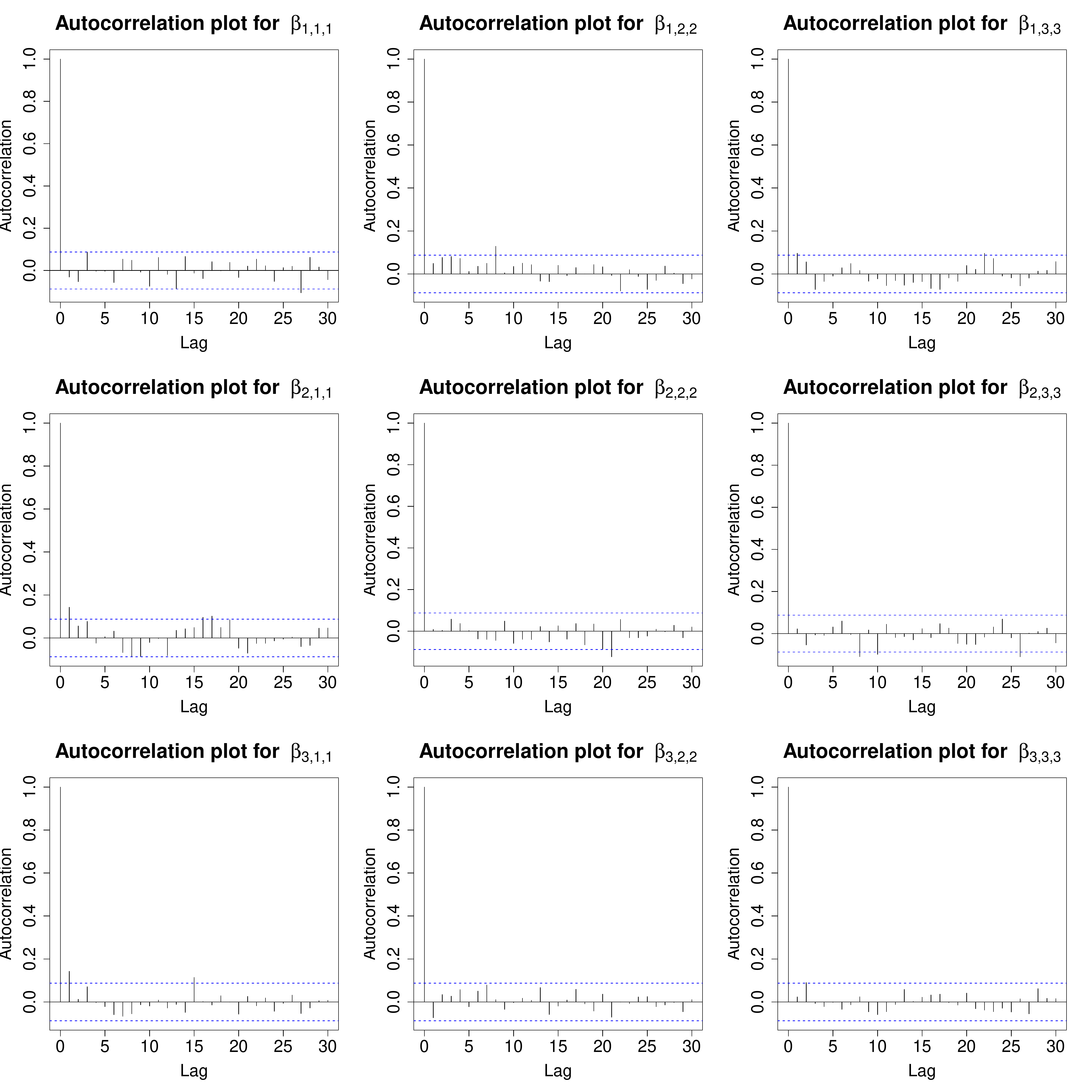} 
    \caption*{Figure S2: Autocorrelation plots for randomly selected $\tbeta_{kqs}$'s using the post-burn-in MCMC samples of the simulation dataset, $k=1,\dots,r_n^o$, $q=1,\dots,Q$, $s=1,\dots,S$. The plots show no signs of non-convergence of the chains.}
\end{figure}

% Figure S3: trace plot 
\begin{figure}[!htb]
    \centering
    \includegraphics[width=\columnwidth]{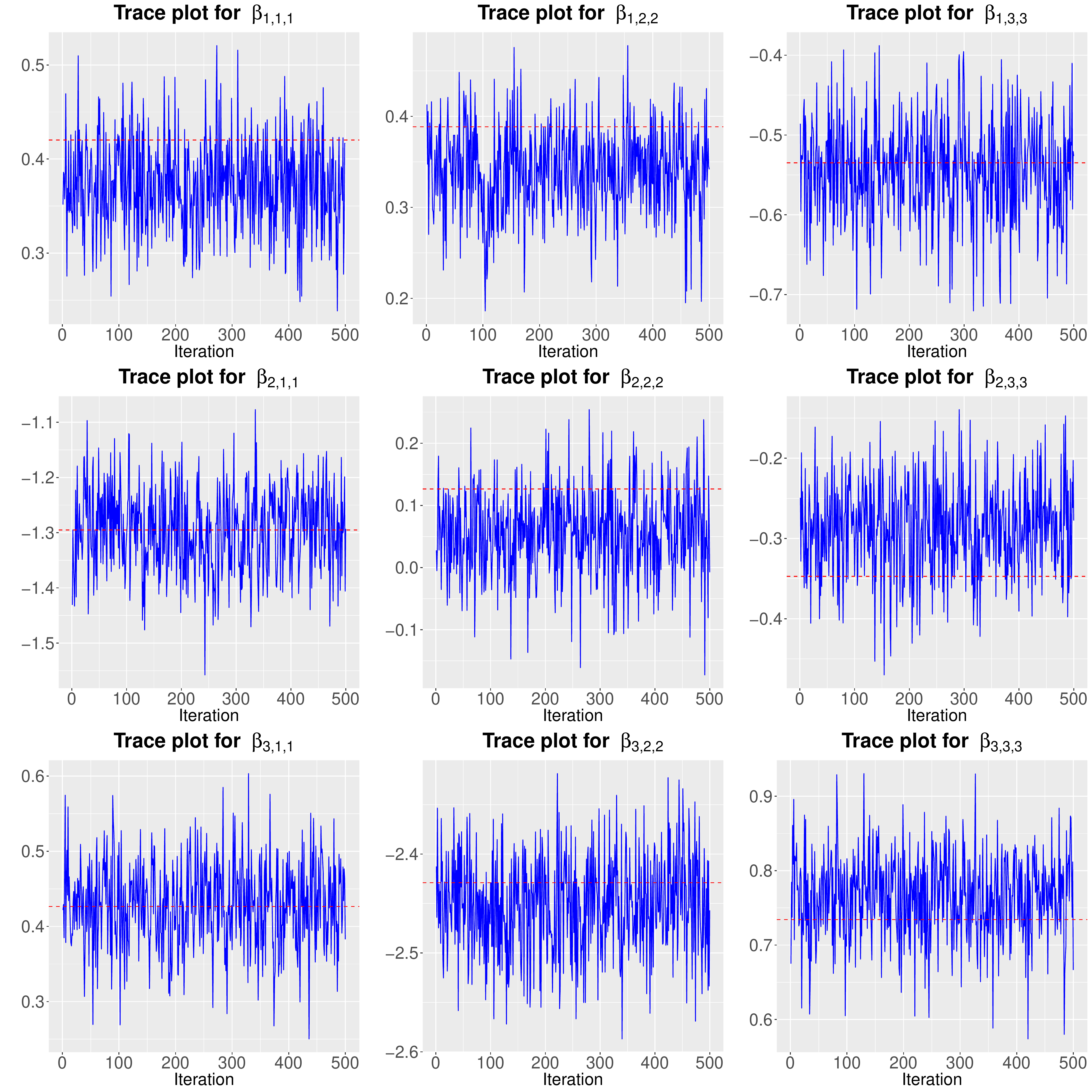} 
    \caption*{Figure S3: Trace plots for randomly selected $\tbeta_{kqs}$'s using the post-burn-in MCMC samples of the simulation dataset, $k=1,\dots,r_n^o$, $q=1,\dots,Q$, $s=1,\dots,S$, where the dashed red lines denote the simulated truths. The plots show no signs of non-convergence of the chains.}
\end{figure}

% Figure S4: combination effect density plot
\begin{figure}[!htb]
    \centering
    \includegraphics[width=\columnwidth]{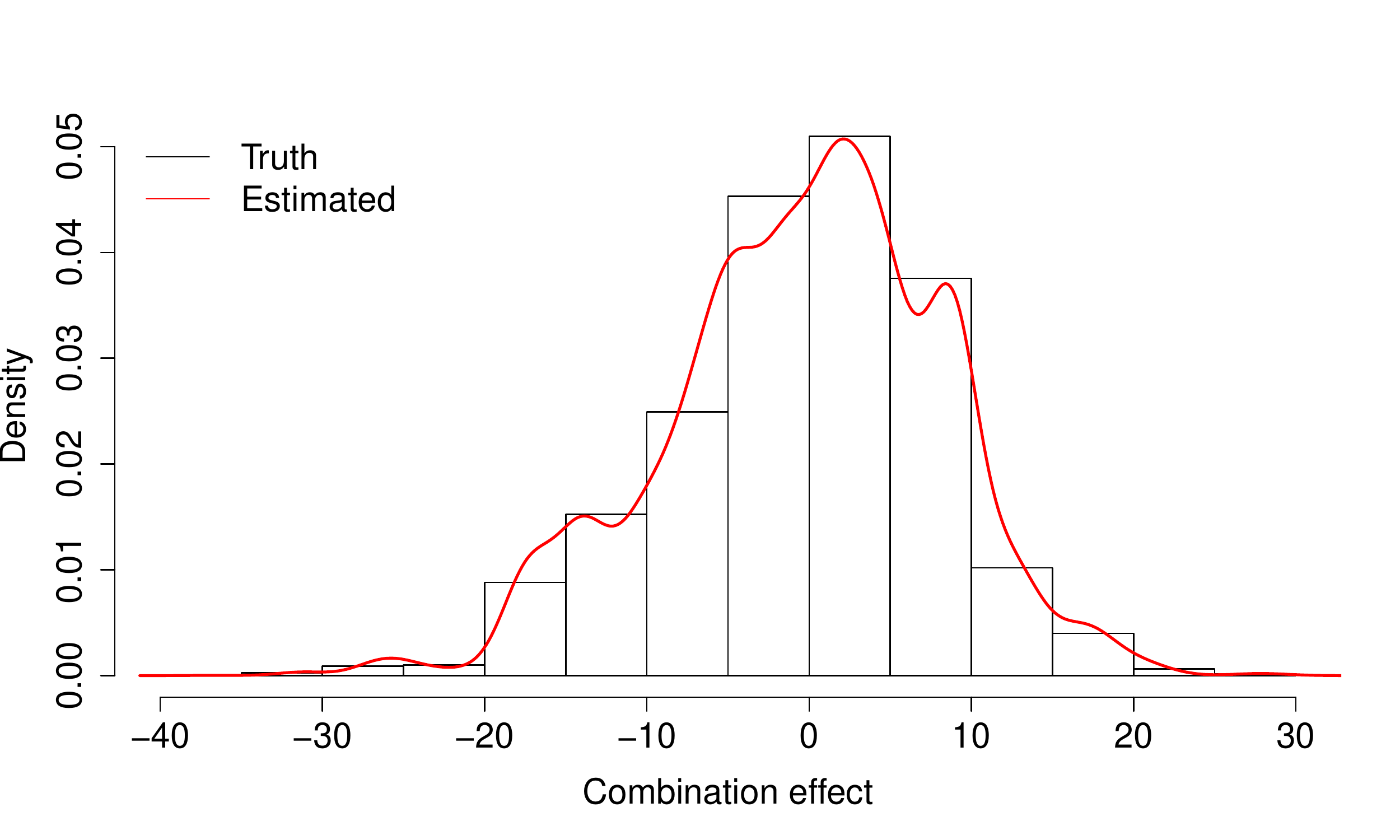} 
    \caption*{Figure S4: Histogram of the true combination effects, overlaid with the empirical density plot of the posterior expected combination effects for one randomly selected simulated dataset in the simulation study. }
\end{figure}

% Table S5: MSE 
\begin{table}[!htb]
\centering
\begin{tabular}{c|c|c}
  \hline
  Parameters & Mean of MSE & Standard Deviation of MSE \\ 
  \hline
  $\btbeta_{11}$ & (5.865e-03, 5.200e-03, 5.788e-03) & (7.327e-03, 3.973e-03, 4.057e-03) \\
  $\btbeta_{12}$ & (5.282e-03, 5.991e-03, 6.080e-03) & (3.893e-03, 4.723e-03, 4.131e-03) \\
  $\btbeta_{13}$ & (6.200e-03, 5.040e-03, 6.234e-03) & (1.196e-02, 4.096e-03, 4.204e-03) \\
  
  $\btbeta_{21}$ & (1.127e-02, 5.459e-03, 7.090e-03) & (9.199e-03, 3.930e-03, 5.213e-03) \\
  $\btbeta_{22}$ & (1.165e-02, 5.450e-03, 7.238e-03) & (9.476e-03, 3.069e-03, 6.012e-03) \\
  $\btbeta_{23}$ & (1.052e-02, 6.386e-03, 6.545e-03) & (7.392e-03, 4.893e-03, 4.385e-03) \\
  
  $\btbeta_{31}$ & (7.978e-03, 5.433e-03, 6.070e-03) & (6.181e-03, 4.939e-03, 4.510e-03) \\
  $\btbeta_{32}$ & (5.966e-03, 5.525e-03, 6.416e-03) & (3.620e-03, 4.622e-03, 5.382e-03) \\
  $\btbeta_{33}$ & (6.657e-03, 5.137e-03, 5.586e-03) & (5.232e-03, 3.908e-03, 3.403e-03) \\
  \hline
\end{tabular}
\caption*{Table S5: Mean and standard deviation of mean squared error (MSE) across 100 simulated datasets for $\{\btbeta_{kq}\}_{k=1}^{r_n^o},_{q=1}^{Q}$. Each entry within the parentheses corresponds to one cluster and one depression item.}
\end{table}

% Figure S6: 95% credible intervals
\begin{figure}[!htb]
    \centering
    \begin{tabular}{ccc}
    \includegraphics[width=.5\textwidth]{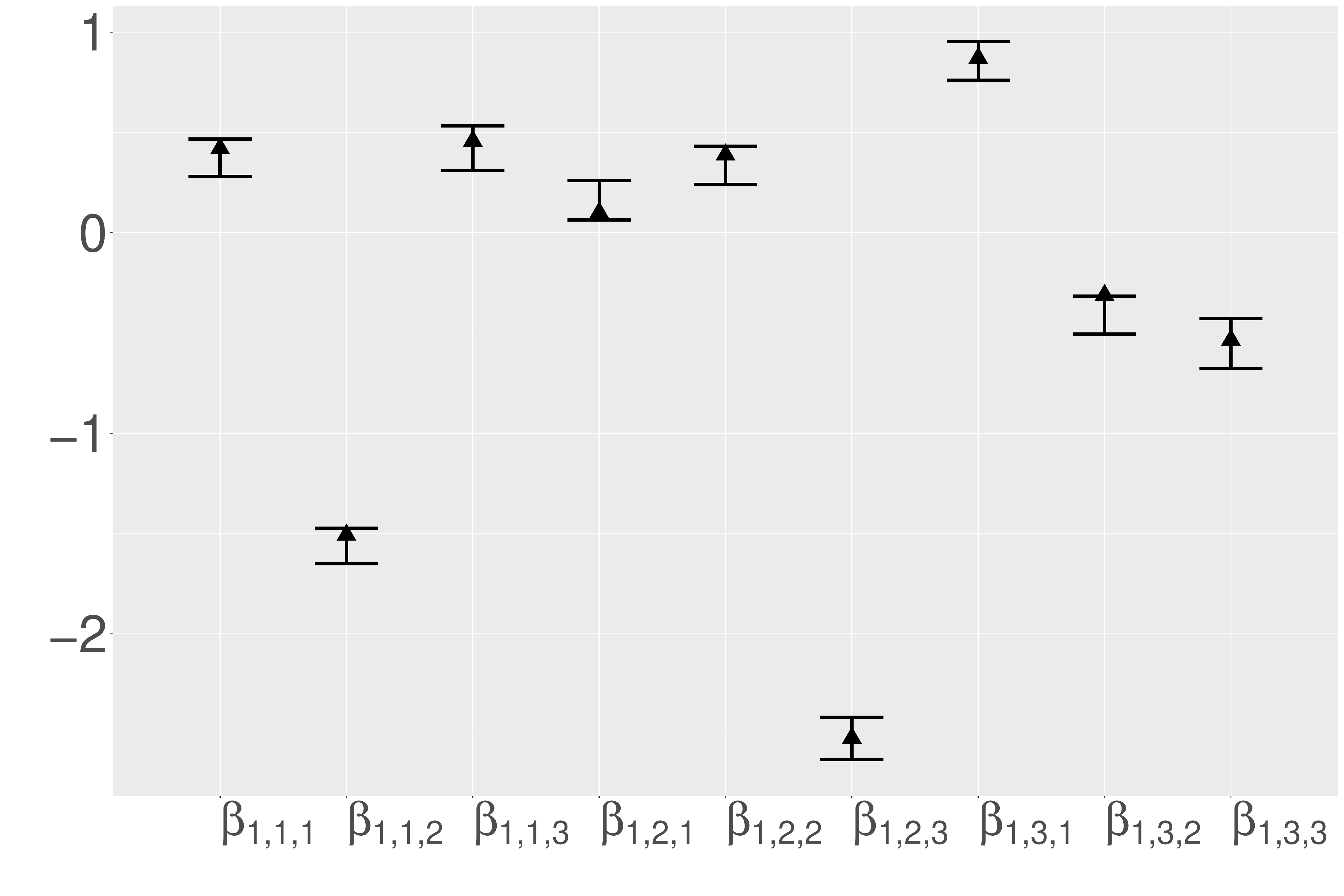} & 
    \includegraphics[width=.5\textwidth]{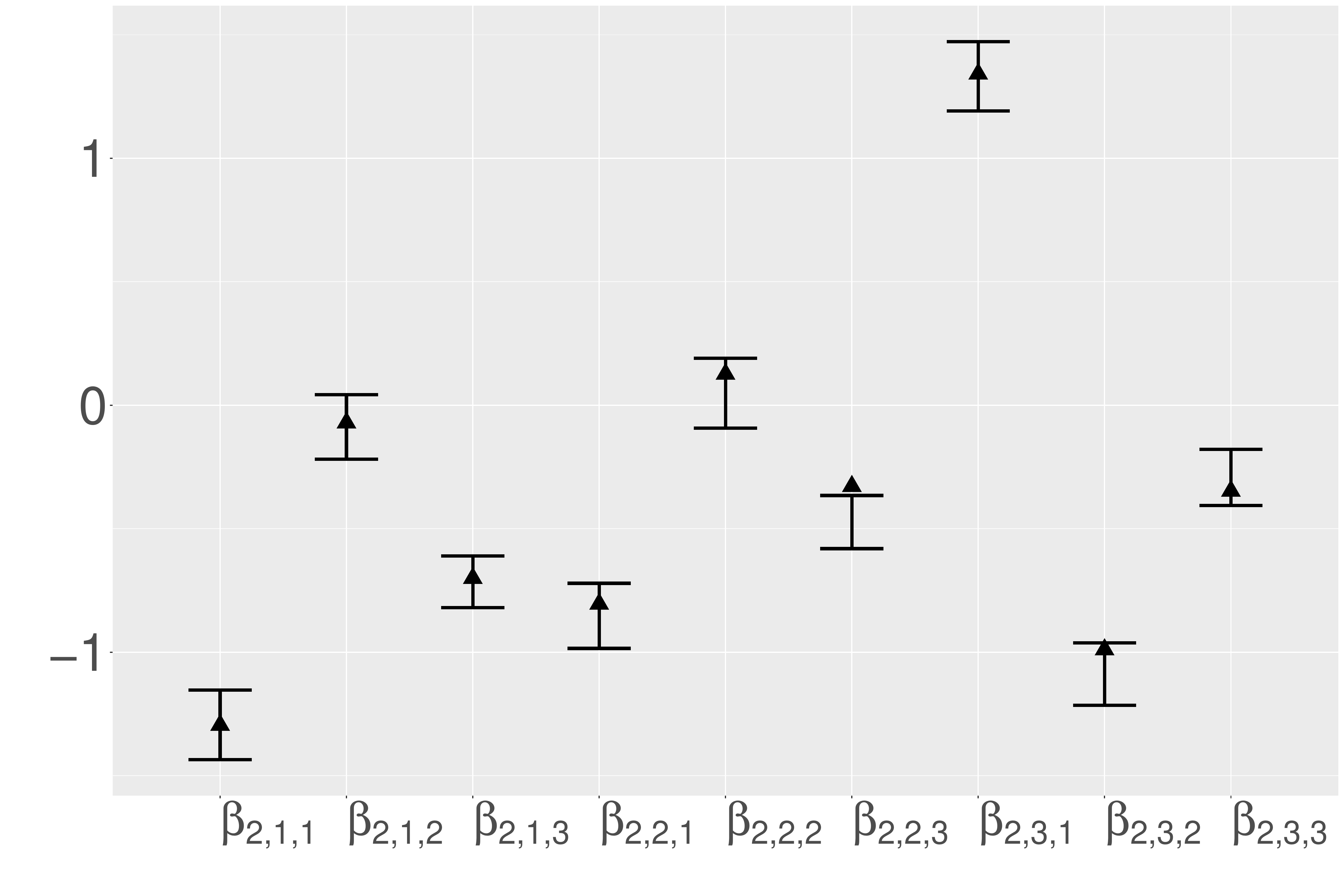} \\
    (a) Coefficients in individual cluster $\#$1 & (b) Coefficients in individual cluster $\#$2 \\ 
    \includegraphics[width=.5\textwidth]{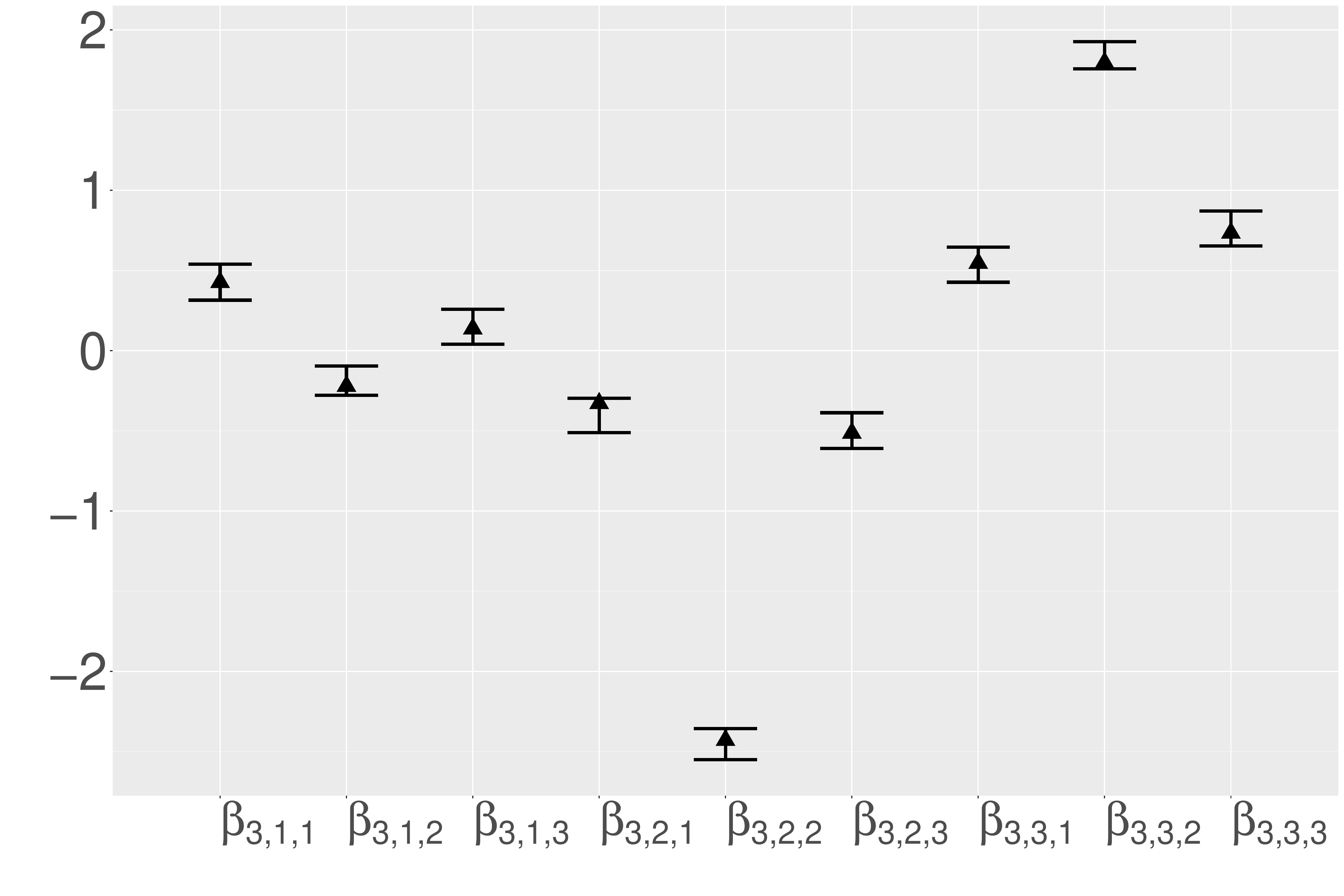} & \\
    (c) Coefficients in individual cluster $\#$3 & \\
    % \multicolumn{2}{c}{\includegraphics[width=.5\textwidth]{./figs/simu_results_c.pdf}} \\
    % \multicolumn{2}{c}{(c) Coefficients in individual cluster $\#$3}
    \end{tabular}
    \caption*{Figure S6: $95\%$ credible intervals of the estimated parameters $\{\tbeta_{kqs}\}_{k=1}^{r_n^o},_{q=1}^{Q},_{s=1}^{S}$  for one randomly selected simulated dataset in the simulation study, where the triangles represent the simulated true values.}
\end{figure}

% Table S7: covariates 
\begin{table}[!htb]
\centering
\begin{adjustbox}{width=\columnwidth*5/6,center}
\begin{tabular}{ p{6cm} p{2cm} p{2cm} p{2cm} p{2cm}}
 \hline
 &  Overall & \multicolumn{3}{c}{Cluster} \\ \cline{3-5}
 &  ($n=259$) & 1($n=132$) & 2($n=84$) & 3($n=43$) \\
 Variables & $n(\%)$ & $n(\%)$ & $n(\%)$ & $n(\%)$ \\ 
 \hline
 \rowcolor{lightgray} \multicolumn{2}{l}{Demographics} & \multicolumn{3}{c}{} \\
 ~\\
 Age &  \\
 \qquad $ <= 25$  &  16(6)   & 8(6)   & 6(7)   & 2(5)    \\
 \qquad $26 - 35$ &  95(37) & 37(28) & 43(51) & 15(35)  \\ 
 \qquad $36 - 45$ &  101(39) & 60(45) & 23(27) & 18(42) \\ 
 \qquad $46 - 55$ &  42(16) & 25(19)  & 11(13) & 6(14)  \\ 
 \qquad $> 55$    &  5(2)   & 2(2)   & 1(1)   & 2(5)  \\ 
 ~\\
 Race &  \\
 \qquad White            & 42(16) & 18(14) & 20(24) & 4(9) \\
 \qquad African-American & 200(77) & 106(80) & 57(68) & 37(86) \\
 \qquad Others           & 17(7) & 8(6)  & 7(8) & 2(5)  \\
 ~\\
 \rowcolor{lightgray} \multicolumn{2}{l}{Clinical Characteristics} & \multicolumn{3}{c}{} \\
 ~\\
 Body Mass Index &  \\
 \qquad $<18.5$       & 4(2)   & 3(2) & 1(1)   & 0(0)  \\
 \qquad $18.5 - 24.9$ & 75(29) & 39(30) & 23(27) & 13(30) \\
 \qquad $25.0 - 29.9$ & 89(34) & 45(34) & 32(38) & 12(28)  \\
 \qquad $\geq 30.0$   & 91(35) & 45(34) & 28(33) & 18(42) \\
 ~\\
 CD4 Count & \\
 \qquad $<=250$    & 69(27) & 38(29) & 17(20) & 14(33)  \\
 \qquad $251-500$  & 109(42) & 58(44) & 39(46) & 12(28)  \\
 \qquad $501-1000$ & 70(27) & 32(24) & 22(26) & 16(37)  \\
 \qquad $>=1001$   & 11(4)  & 4(3)   & 6(7)  & 1(2)  \\
 ~\\
 Viral Load &  \\
 \qquad $<=500$      & 136(53) & 70(53) & 45(54) & 21(49) \\
 \qquad $501-5000$   & 41(16)  & 17(13)  & 16(19) & 8(19)  \\
 \qquad $5001-50000$ & 50(19) & 29(22) & 12(14) & 9(21) \\
 \qquad $>=50001$    & 32(12)  & 16(12) & 11(13) & 5(12)   \\
 ~\\
 Hypertension & \\
 \qquad Yes & 66(25) & 36(27) & 18(21) & 12(28) \\
 \qquad No  & 193(75) & 96(73) & 66(79) & 31(72)  \\
 ~\\
 \rowcolor{lightgray} \multicolumn{2}{l}{Behavioral Characteristics} & \multicolumn{3}{c}{} \\
 ~\\
 Smoke Status & \\
 \qquad Yes & 110(42) & 51(39) & 37(44) & 22(51) \\
 \qquad No  & 149(58) & 81(61) & 47(56) & 21(49) \\
 ~\\
 Substance Use & \\
 \qquad Yes & 38(15) & 20(15) & 12(14) & 6(14) \\
 \qquad No  & 221(85) & 112(85) & 72(86) & 37(86)  \\
 \hline
\end{tabular}
\end{adjustbox}
\caption*{Table S7: Demographic, clinical and behavioral characteristics of individuals at the initial visit in the overall sample and in the three clusters.}
\end{table}

% Table S8: ART drugs 
\begin{table}[!htb]
\centering
\begin{adjustbox}{width=\columnwidth*6/7,center}
\begin{tabular}{ p{7cm} p{2cm} p{2cm} p{2cm} p{2cm} }
  \hline
  &  Overall & \multicolumn{3}{c}{Cluster} \\ \cline{3-5}
  &  ($n=259$) & 1($n=132$) & 2($n=84$) & 3($n=43$)   \\
  ART Drugs & $n(\%)$ & $n(\%)$ & $n(\%)$ & $n(\%)$ \\ 
 \hline
 \rowcolor{lightgray} NRTI & \multicolumn{4}{c}{}  \\
 Abacavir (ABC)       & 109(42) & 60(45) & 29(35) & 20(47)  \\
 Zidovudine (AZT)    & 166(64) & 79(60) & 59(70) & 28(65)  \\
 Stavudine (D4T)      & 115(44) & 57(43) & 36(43) & 22(51) \\
 Zalcitabine (DDC)    & 25(10)  & 13(10) & 10(12) & 2(5) \\ 
 Didanosine (DDI)     & 91(35) & 44(33) & 35(42) & 12(28) \\ 
 Emtricitabine (FTC)  & 192(74) & 101(77) & 60(71) & 31(72) \\
 Lamivudine (LAM)     & 213(82) & 105(80) & 70(83) & 38(88) \\
 Tenofovir Disoproxil Fumarate (TDF) & 210(81) & 110(83) & 67(80) & 33(77) \\
 ~\\
 \rowcolor{lightgray} NNRTI & \multicolumn{4}{c}{}  \\
 Efavirenz (EFV)  & 128(49) & 58(44) & 46(55) & 24(56)  \\
 Etravirine (ETV) &  20(8)  &  14(11)  & 1(1)   & 5(12)   \\
 Nevirapine (NVP) & 64(25) & 33(25) & 25(30) & 6(14)  \\
 Rilpivirine (RPV) & 23(9) & 13(10) & 7(8)  & 3(7)   \\

 ~\\
 \rowcolor{lightgray} PI & \multicolumn{4}{c}{}  \\
 Atazanavir (ATZ)    & 102(39) & 50(38) & 33(39) & 19(44) \\
 Darunavir (DRV)     & 52(20) & 27(20)  & 11(13) &  14(33)  \\
 Fosamprenavir (FPV) &  22(8)  & 13(10)   &  6(7)  &  3(7)   \\ 
 Indinavir (IDV)     & 63(24) & 29(22) & 24(29) & 10(23) \\
 Lopinavir (LPV)     & 73(28) & 37(28) & 26(31) & 10(23)    \\
 Nelfinavir (NFV)    & 75(29) & 33(25) & 28(33) & 14(33)  \\
 Ritonavir (RTV)     & 135(52) & 68(52) & 41(49) & 26(60)  \\
 Saquinavir (SQV)    & 32(12) & 14(11) & 11(13) & 7(16)  \\
 ~\\
 \rowcolor{lightgray} INSTI & \multicolumn{4}{c}{}  \\
 Dolutegravir (DGT)   & 35(14) & 19(14) & 10(12) & 6(14) \\
 Elvitegravir (ELV)   & 15(6)  & 10(8)   & 4(5) & 1(2)    \\
 Raltegravir (RAL)    & 55(21) & 35(27) & 12(14) & 8(19)  \\
 ~\\
\rowcolor{lightgray} EI & \multicolumn{4}{c}{}  \\
 Maraviroc (SLZ)  & 5(2) & 3(2) & 0(0) & 2(5) \\
 \hline
\end{tabular}
\end{adjustbox}
\caption*{Table S8: ART drugs used by individuals in the overall sample and in the three clusters, where NRTI denotes nucleoside reverse-transcriptase inhibitors, NNRTI denotes non-nucleoside reverse-transcriptase inhibitor, PI denotes protease inhibitor, INSTI denotes integrase inhibitor, and EI denotes entry inhibitor.}
\end{table}

% Figure S9: Rshiny 
\begin{figure}[!htb]
    \centering
    \includegraphics[width=\columnwidth]{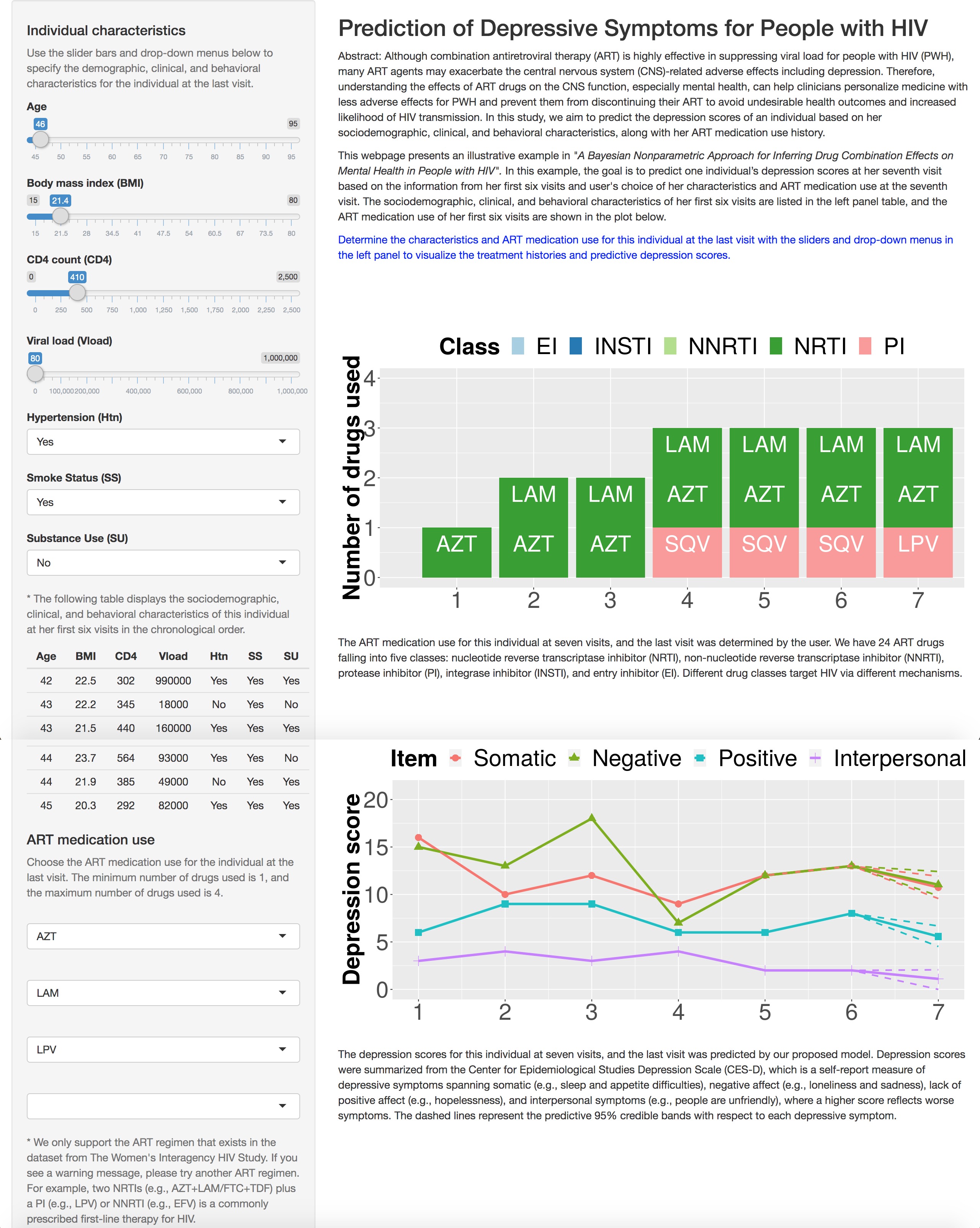} 
    \caption*{Figure S9: A screenshot of the R Shiny web application. The web user interface interactively displays the predictive depression scores of an individual in response to the user's choice of the individual's clinical characteristics and ART medication use. }
\end{figure}

\end{document}